%% file: main.tex
\documentclass[twocolumn, tighten, dvipsnames]{aastex63}
\received{2021 February 15}
\revised{2021 May 20}
\accepted{2021 June 9}

\submitjournal{ApJ}

\shorttitle{MAPS IX: The Large Organic Molecules \ce{HC3N}, \ce{CH3CN}, and $c$-\ce{C3H2}}
\shortauthors{Ilee et al.}

\usepackage{xspace}
\usepackage[version=4]{mhchem}

\def\arraystretch{1.3}

\begin{document}

\title{Molecules with ALMA at Planet-forming Scales (MAPS) IX: \\ Distribution and properties of the large organic molecules \ce{HC3N}, \ce{CH3CN}, and \boldmath{$c$}-\ce{C3H2}}

\correspondingauthor{John D. Ilee}
\email{j.d.ilee@leeds.ac.uk}

\author[0000-0003-1008-1142]{John~D.~Ilee} 
\affiliation{School of Physics and Astronomy, University of Leeds, Leeds, UK, LS2 9JT}

\author[0000-0001-6078-786X]{Catherine~Walsh}
\affiliation{School of Physics and Astronomy, University of Leeds, Leeds, UK, LS2 9JT}

\author[0000-0003-2014-2121]{Alice~S.~Booth}
\affiliation{Leiden Observatory, Leiden University, 2300 RA Leiden, the Netherlands}
\affiliation{School of Physics and Astronomy, University of Leeds, Leeds, UK, LS2 9JT}

\author[0000-0003-3283-6884]{Yuri Aikawa}
\affiliation{Department of Astronomy, Graduate School of Science, The University of Tokyo, Tokyo 113-0033, Japan}

\author[0000-0003-2253-2270]{Sean M. Andrews}
\affiliation{Center for Astrophysics \textbar\, Harvard \& Smithsonian, 60 Garden St., Cambridge, MA 02138, USA}

\author[0000-0001-7258-770X]{Jaehan Bae}
\altaffiliation{NASA Hubble Fellowship Program Sagan Fellow}
\affiliation{Department of Astronomy, University of Florida, Gainesville, FL 32611, USA}
\affiliation{Earth and Planets Laboratory, Carnegie Institution for Science, 5241 Broad Branch Road NW, Washington, DC 20015, USA}

\author[0000-0003-4179-6394]{Edwin A.\ Bergin}
\affiliation{Department of Astronomy, University of Michigan, 323 West Hall, 1085 South University Avenue, Ann Arbor, MI 48109, USA}

\author[0000-0002-8716-0482]{Jennifer B. Bergner}
\altaffiliation{NASA Hubble Fellowship Program Sagan Fellow}
\affiliation{University of Chicago Department of the Geophysical Sciences, Chicago, IL 60637, USA}

\author[0000-0003-4001-3589]{Arthur D. Bosman}
\affiliation{Department of Astronomy, University of Michigan, 323 West Hall, 1085 South University Avenue, Ann Arbor, MI 48109, USA}

\author[0000-0002-2700-9676]{Gianni Cataldi}
\affiliation{National Astronomical Observatory of Japan, 2-21-1 Osawa, Mitaka, Tokyo 181-8588, Japan}
\affiliation{Department of Astronomy, Graduate School of Science, The University of Tokyo, Tokyo 113-0033, Japan}

\author[0000-0003-2076-8001]{L. Ilsedore Cleeves}
\affiliation{Department of Astronomy, University of Virginia, Charlottesville, VA 22904, USA}

\author[0000-0002-1483-8811]{Ian Czekala}
\altaffiliation{NASA Hubble Fellowship Program Sagan Fellow}
\affiliation{Department of Astronomy \& Astrophysics, 525 Davey Laboratory, The Pennsylvania State University, University Park, PA 16802, USA}
\affiliation{Center for Exoplanets \& Habitable Worlds, 525 Davey Laboratory, The Pennsylvania State University, University Park, PA 16802, USA}
\affiliation{Center for Astrostatistics, 525 Davey Laboratory, The Pennsylvania State University, University Park, PA 16802, USA}
\affiliation{Institute for Computational \& Data Sciences, The Pennsylvania State University, University Park, PA 16802, USA}
\affiliation{Department of Astronomy, 501 Campbell Hall, University of California, Berkeley, CA 94720-3411, USA}

\author[0000-0003-4784-3040]{Viviana V. Guzm\'{a}n}
\affiliation{Instituto de Astrof\'isica, Pontfificia Universidad Cat\'olica de Chile, Av. Vicu\~na Mackenna 4860, 7820436 Macul, Santiago, Chile}

\author[0000-0001-6947-6072]{Jane Huang}
\altaffiliation{NASA Hubble Fellowship Program Sagan Fellow}
\affiliation{Department of Astronomy, University of Michigan, 323 West Hall, 1085 South University Avenue, Ann Arbor, MI 48109, USA}
\affiliation{Center for Astrophysics \textbar\, Harvard \& Smithsonian, 60 Garden St., Cambridge, MA 02138, USA}

\author[0000-0003-1413-1776]{Charles J. Law}
\affiliation{Center for Astrophysics \textbar\, Harvard \& Smithsonian, 60 Garden St., Cambridge, MA 02138, USA}

\author[0000-0003-1837-3772]{Romane Le Gal}
\affiliation{IRAP, Universit\'{e} de Toulouse, CNRS, CNES, UT3, 31400 Toulouse, France}
\affiliation{Univ. Grenoble Alpes, CNRS, IPAG, F-38000 Grenoble, France}
\affiliation{IRAM, 300 rue de la piscine, F-38406 Saint-Martin d'H\`{e}res, France}
\affiliation{Center for Astrophysics \textbar\, Harvard \& Smithsonian, 60 Garden St., Cambridge, MA 02138, USA}

\author[0000-0002-8932-1219]{Ryan A. Loomis}
\affiliation{National Radio Astronomy Observatory, 520 Edgemont Rd., Charlottesville, VA 22903, USA}

\author[0000-0002-1637-7393]{Fran\c cois M\'enard}
\affiliation{Univ. Grenoble Alpes, CNRS, IPAG, F-38000 Grenoble, France}

\author[0000-0002-7058-7682]{Hideko Nomura}
\affiliation{National Astronomical Observatory of Japan, 2-21-1 Osawa, Mitaka, Tokyo 181-8588, Japan}

\author[0000-0001-8798-1347]{Karin I. \"Oberg}
\affiliation{Center for Astrophysics \textbar\, Harvard \& Smithsonian, 60 Garden St., Cambridge, MA 02138, USA}

\author[0000-0001-8642-1786]{Chunhua Qi}
\affiliation{Center for Astrophysics \textbar\, Harvard \& Smithsonian, 60 Garden St., Cambridge, MA 02138, USA}

\author[0000-0002-6429-9457]{Kamber R. Schwarz}
\altaffiliation{NASA Hubble Fellowship Program Sagan Fellow}
\affiliation{Lunar and Planetary Laboratory, University of Arizona, 1629 E. University Blvd, Tucson, AZ 85721, USA}

\author[0000-0003-1534-5186]{Richard Teague}
\affiliation{Center for Astrophysics \textbar\, Harvard \& Smithsonian, 60 Garden St., Cambridge, MA 02138, USA}

\author[0000-0002-6034-2892]{Takashi Tsukagoshi} 
\affiliation{National Astronomical Observatory of Japan, 2-21-1 Osawa, Mitaka, Tokyo 181-8588, Japan}

\author[0000-0003-1526-7587]{David J. Wilner}
\affiliation{Center for Astrophysics \textbar\, Harvard \& Smithsonian, 60 Garden St., Cambridge, MA 02138, USA}

\author[0000-0003-4099-6941]{Yoshihide Yamato}
\affiliation{Department of Astronomy, Graduate School of Science, The University of Tokyo, Tokyo 113-0033, Japan}

\author[0000-0002-0661-7517]{Ke Zhang}
\altaffiliation{NASA Hubble Fellow}
\affiliation{Department of Astronomy, University of Wisconsin-Madison, 475 N Charter St, Madison, WI 53706}
\affiliation{Department of Astronomy, University of Michigan, 323 West Hall, 1085 South University Avenue, Ann Arbor, MI 48109, USA}

\begin{abstract}
The precursors to larger, biologically-relevant molecules are detected throughout interstellar space, but determining the presence and properties of these molecules during planet formation requires observations of protoplanetary disks at high angular resolution and sensitivity.  Here we present $0\farcs3$ observations of  \ce{HC3N}, \ce{CH3CN}, and $c$-\ce{C3H2} in five protoplanetary disks observed as part of the Molecules with ALMA at Planet-forming Scales (MAPS) Large Program.  We robustly detect all molecules in four of the disks (GM~Aur, AS~209, HD~163296 and MWC~480) with tentative detections of $c$-\ce{C3H2} and \ce{CH3CN} in IM Lup.  We observe a range of morphologies -- central peaks, single or double rings -- with no clear correlation in morphology between molecule nor disk.  Emission is generally compact and on scales comparable with the millimetre dust continuum.  We perform both disk-integrated and radially-resolved rotational diagram analysis to derive column densities and rotational temperatures.  The latter reveals 5--10 times more column density in the inner 50--100~au of the disks when compared with the disk-integrated analysis.  We demonstrate that \ce{CH3CN} originates from lower relative heights in the disks when compared with \ce{HC3N}, in some cases directly tracing the disk midplane.  Finally, we find good agreement between the ratio of small to large nitriles in the {outer disks} and comets. Our results indicate that the protoplanetary disks studied here are host to significant reservoirs of large organic molecules, and that this planet- and comet-building material {can be} chemically similar to that in our own Solar System. This paper is part of the MAPS special issue of the Astrophysical Journal Supplement Series.
\end{abstract}

%\keywords{protoplanetary disks -- planet formation -- comets -- astrochemistry --- interstellar molecules}


\section{Introduction} 
\label{sec:intro}

Protoplanetary disks host the basic ingredients for planet formation. 
The abundance and spatial distribution of organic molecules are of particular importance as the most complex of these --- the so-called \textit{complex organic molecules} (COMs) --- are the precursors of larger, pre-biotic, molecules \citep[see, e.g.,][]{Herbst2009}.  
These molecules are the vital bridges between the relatively simple molecules that are abundant in circumstellar environments, e.g. \ce{CO}, and those important for life, such as amino acids. 
We know that the surface of the young Earth was seeded with organic material via impacts from planetesimals (comets and asteroids) that formed within the disk around the young Sun \citep[see][]{altwegg_2019}. However, it remains unclear whether or not a complex organic reservoir is present in all protoplanetary disks, thus determining their propensity for developing life-friendly environments.

\begin{figure*}[!ht]
\centering
\includegraphics[width=\linewidth, trim=0 0.5cm 0 1.5cm, clip]{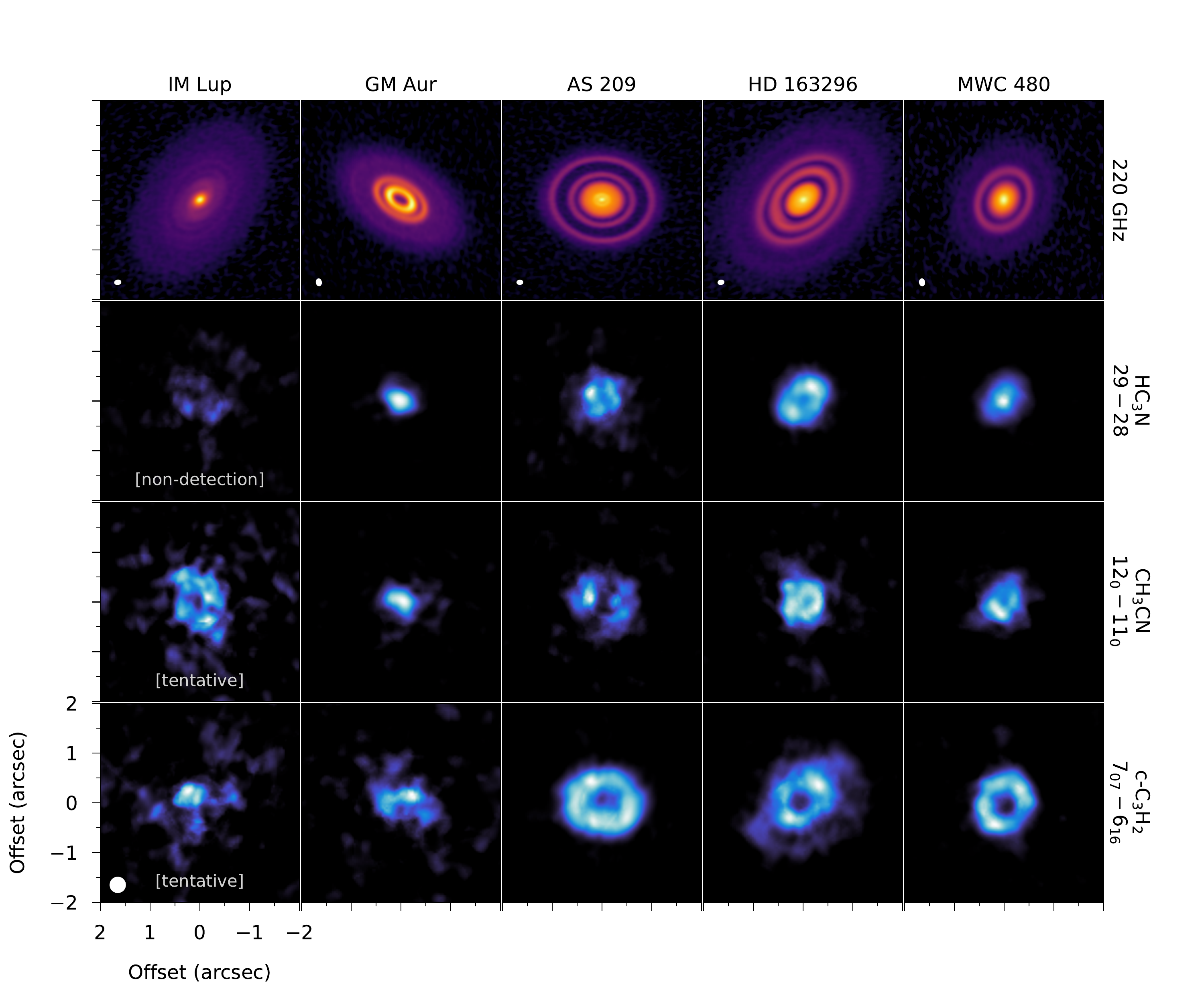}
\caption{Continuum images at 220\,GHz (top) shown alongside integrated intensity (zeroth moment) maps for the brightest transitions of \ce{HC3N}, \ce{CH3CN}, and $c$-\ce{C3H2} in our target disks (bottom).  Panels are normalised with continuum and integrated intensity images displayed with power-law and linear stretches, respectively. Beams are shown as a filled ellipse (which is identical for all integrated intensity maps).  {See Appendix \ref{sec:mom0} for a gallery of all transitions with intensity scales}, and \citet{sierra20} for further information on the continuum images.}
\label{fig:cont_and_best}
\end{figure*}

The study of the chemical content of protoplanetary disks began during the advent of (sub)millimetre astronomy that enabled the detection of rotational transitions of small molecules such as \ce{CO}, \ce{HCO+}, \ce{CN}, \ce{CS}, \ce{C2H}, \ce{HCN}, \ce{HNC}, and \ce{H2CO} \citep[e.g.,][]{Dutrey1997,Kastner1997,vanZadelhoff2001,Aikawa2003,Thi2004,Oberg2010}.
In these pioneering studies, it was realised that the gas-phase abundances of these species in protoplanetary disks were orders of magnitude lower than those in nearby dark clouds. 
The prevailing explanation is that disks have a cold dense midplane where most species are frozen out as ices on dust grain surfaces, and a surface layer where dissociation and ionization dominates.   Gas-phase molecules are confined to a warm molecular layer between these two regions \citep{Aikawa2002}.
Models suggest that larger molecules can be efficiently formed within the ices on dust grains \citep[e.g.,][]{Walsh2014}, and that processes such as non-thermal desorption are required to release these strongly-bound molecules into the gas-phase.  Thus, the detection of emission from such species provides insight into the composition and distribution of the organic ice reservoir.

The younger phases ($\lesssim 1$~Myr) of both low- and high-mass star formation host organic-rich reservoirs {(for reviews see \citealt{Herbst2009}, \citealt{Caselli2012} and \citealt{Jorgensen2020})}, but it remains unclear what degree of this organic-rich material is inherited by the protoplanetary disk \citep[e.g.,][]{Drozdovskaya2016, Drozdovskaya2019, Bianchi2019}. 
The unique chemical structure of protoplanetary disks and their small angular size presents challenges when attempting to detect rotational line emission from larger organic molecules.  
Because the bulk of the mass of protoplanetary disks has a temperature less than $\sim 100$~K, most large organics are hosted on ice mantles, significantly reducing the gas-phase abundances.  
In addition, the larger the molecule, the more complex the spectrum, and the larger the partition function, leading to (in general) significantly weaker emission for individual transitions \citep{Herbst2009}.  
Hence, it has only been very recently that organic molecules with more than four atoms have been successfully detected in protoplanetary disks, and this has been facilitated by the availability of very sensitive, high-angular resolution observations. 
%with interferometers such as SMA, PdBI/NOEMA and most recently, ALMA.

Searches facilitated by interferometers have revealed the prevalence of larger hydrocarbons and complex nitriles in protoplanetary disks.
\citet{Chapillon2012}, \citet{Qi2013}, and \citet{Oberg2015} report the first detections of gas-phase \ce{HC3N} (cyanoacetylene), $c$-\ce{C3H2} (cyclopropanylidene), and \ce{CH3CN} (methyl cyanide) in protoplanetary disks, respectively.   
Follow-up studies and surveys, facilitated by the Atacama Large Millimeter/submillimeter Array (ALMA), have confirmed the relative ubiquity of these molecules in several nearby well-studied protoplanetary disks \citep{Bergin2016,Kastner2018,Bergner2018,Loomis2020,Facchini2021}. 
Their prevalence appears to be connected to an enhanced ratio of elemental carbon relative to elemental oxygen in the disk atmosphere, caused by physical and chemical processing of oxygen-rich ices that effectively removes oxygen from the gas-phase \citep[e.g.,][]{Kama2016,Du2017}.
{On the other hand, oxygen-bearing COMs have remained elusive \citep{Carney2019}.  Until recently, \ce{CH3OH} and \ce{HCOOH} had only been detected in the closest protoplanetary disk, TW~Hya \citep{Walsh2016,Favre2018}.  Further detections of \ce{CH3OH} are now beginning to emerge, but these have so far been confined to young, warm disks \citep{vantHoff2018, Lee2019, Podio2020} or those with irradiated cavities \citep{Booth2021}}.  Nonetheless, models have shown that a combination of both gas-phase chemistry and ice-mantle chemistry is needed to explain the abundances of both the complex nitriles and O-bearing organics detected thus far \citep[][]{Loomis2018a,LeGal2019}.  

\begin{figure*}[!ht]
\centering
\includegraphics[width=\linewidth, clip]{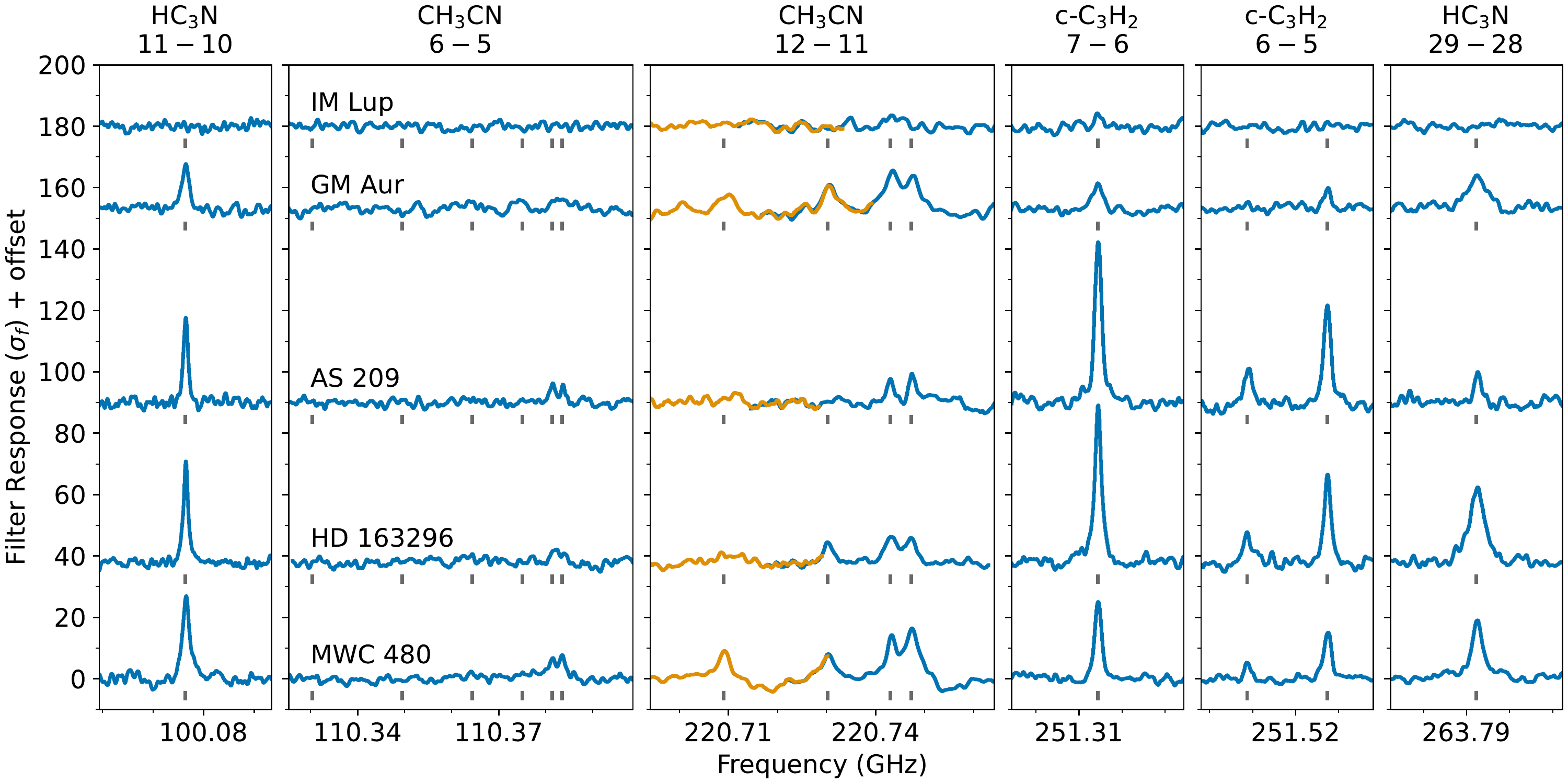}
\caption{Matched filter response $\sigma_{f}$ across the (rest) frequency range each of the targeted transitions (vertically offset for clarity).  Multiple spectral windows are shown with different line colors, and vertical dashed lines mark the rest frequencies of transitions.}
\label{fig:filter_all}
\end{figure*}

In this paper, we report high-angular resolution (0\farcs3) observations of multiple transitions of \ce{HC3N}, $c$-\ce{C3H2}, and \ce{CH3CN} towards five nearby protoplanetary disks with ALMA (IM~Lup, GM~Aur, AS~209, HD~163296, and MWC~480). These data were collected as part of the Molecules with ALMA at Planet-forming Scales (MAPS) Large Program, which has the overarching aim of elucidating the chemistry of planet formation (see \citealt{oberg20} for further details), and the paper is structured as follows.  In Section \ref{sec:methods} we outline our methods for detecting, imaging, and analysing the line emission from each of the target disks.  In Section \ref{sec:results} we report images of the line emission, present azimuthally-averaged radial profiles of the emission, and use these data to determine both disk-averaged and radially-resolved column densities and rotational temperatures (where data quality allows).  We compare the obtained molecular distributions and rotational temperatures with those available in the literature.  In Section \ref{sec:discussion} we discuss the trends present in our sample, and compare our retrieved parameters with available model results to constrain the chemical origin of these larger hydrocarbons and nitriles.  We also compare the radial emission profiles of each molecule with the millimetre dust emission of each disk, and discuss implications of the dust sculpting on the resulting molecular distributions and emission patterns. Finally, we present a future outlook for studies of complex organic molecules in protoplanetary disks. 

\section{Methods}
\label{sec:methods}

\subsection{Overview of observations}
\label{sec:observations}
The data presented here were collected as part of the ALMA Large Program MAPS\footnote{ \url{http://www.alma-maps.info}} (project ID 2018.1.01055.L, co-PIs, K.~I. \"{O}berg, Y.~Aikawa, E.~A. Bergin, V.~V. Guzm\'{a}n, and C.~Walsh).  Full details of the scientific scope and targets of the program are provided in \citet{oberg20}.  Hence, we provide only a brief overview here and limit our description to information pertinent to the scope of this work.

{MAPS targeted five protoplanetary disks -- IM~Lup, GM~Aur, AS~209, HD~163296, and MWC~480 -- using four spectral settings: two in Band 3 and two in Band 6.  Together these settings covered 15 transitions of the large organic molecules \ce{HC3N}, \ce{CH3CN} and $c$-\ce{C3H2}.  Observations using two antenna configurations (compact and extended) were used to recover both large and small scale emission from the disks. Standard calibration routines were initially performed by ALMA staff, supplemented by additional self-calibration to improve the signal-to-noise ratio (see \citet{oberg20} for full details).}

\subsection{Matched filtering}
\label{sec:filter}

Several of our target lines are predicted to be relatively weak; hence, we initially processed the measurement sets of the line-containing spectral windows using a matched filter as described in \citet{Loomis2018b} (see Figure \ref{fig:filter_all}). 
Several Keplerian filters were attempted with varying radial extents, from the full spatial extent of the $^{12}$CO emission, down to very compact filters matched to the emitting radius seen in test images of the target lines.  We used the matched filtering response $\sigma_{f}$ to define two classes of line detection -- tentative (where $3 < \sigma_{f} < 5$), and robust (where $\sigma_{f} > 5$).  This ensured our criteria for detection were not influenced by our choice of imaging parameters.  We further discuss the results of the matched filtering in Section \ref{res:matchedfilter}.

\begin{figure*}[!ht]
\centering
\includegraphics[width=0.32\textwidth]{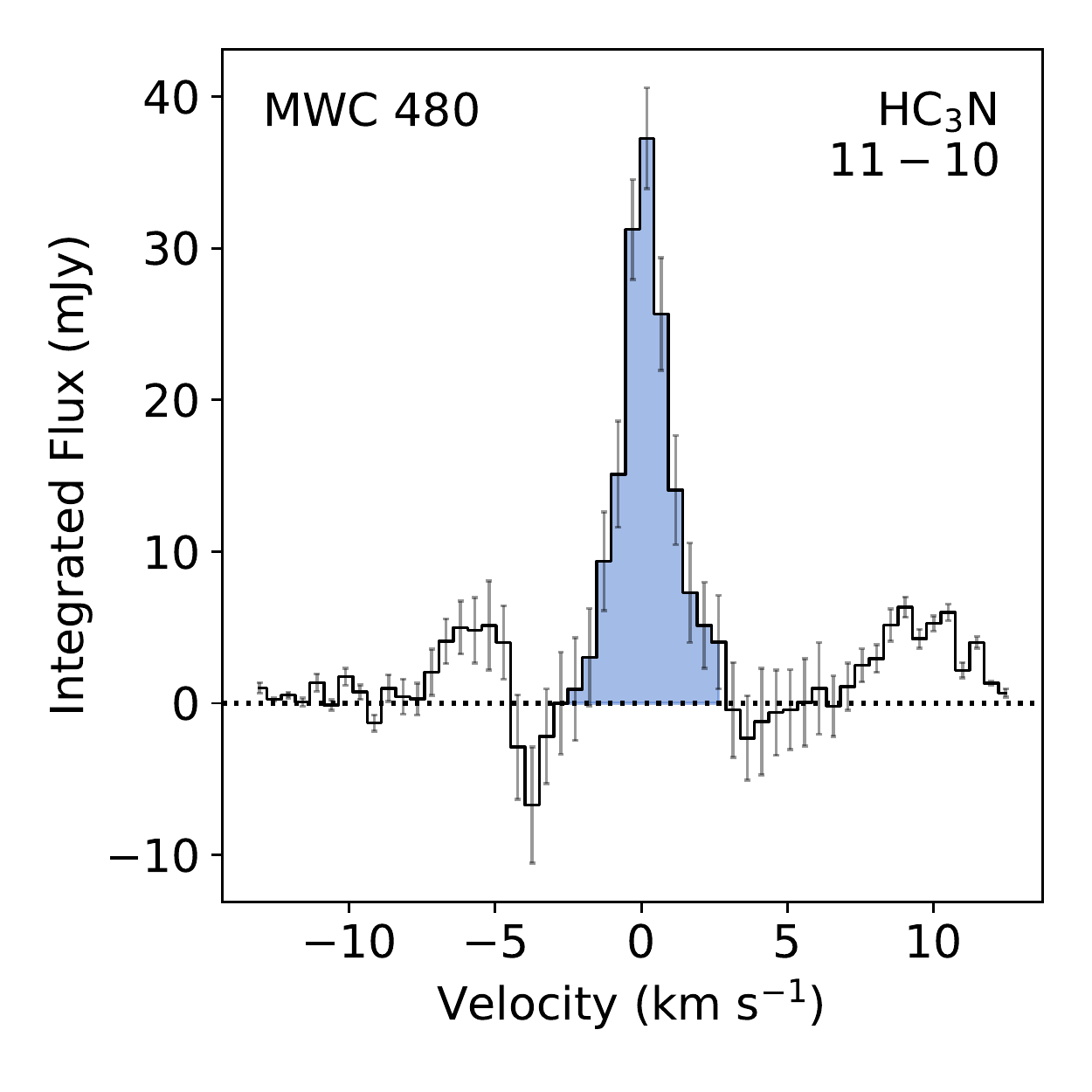}
\includegraphics[width=0.32\textwidth]{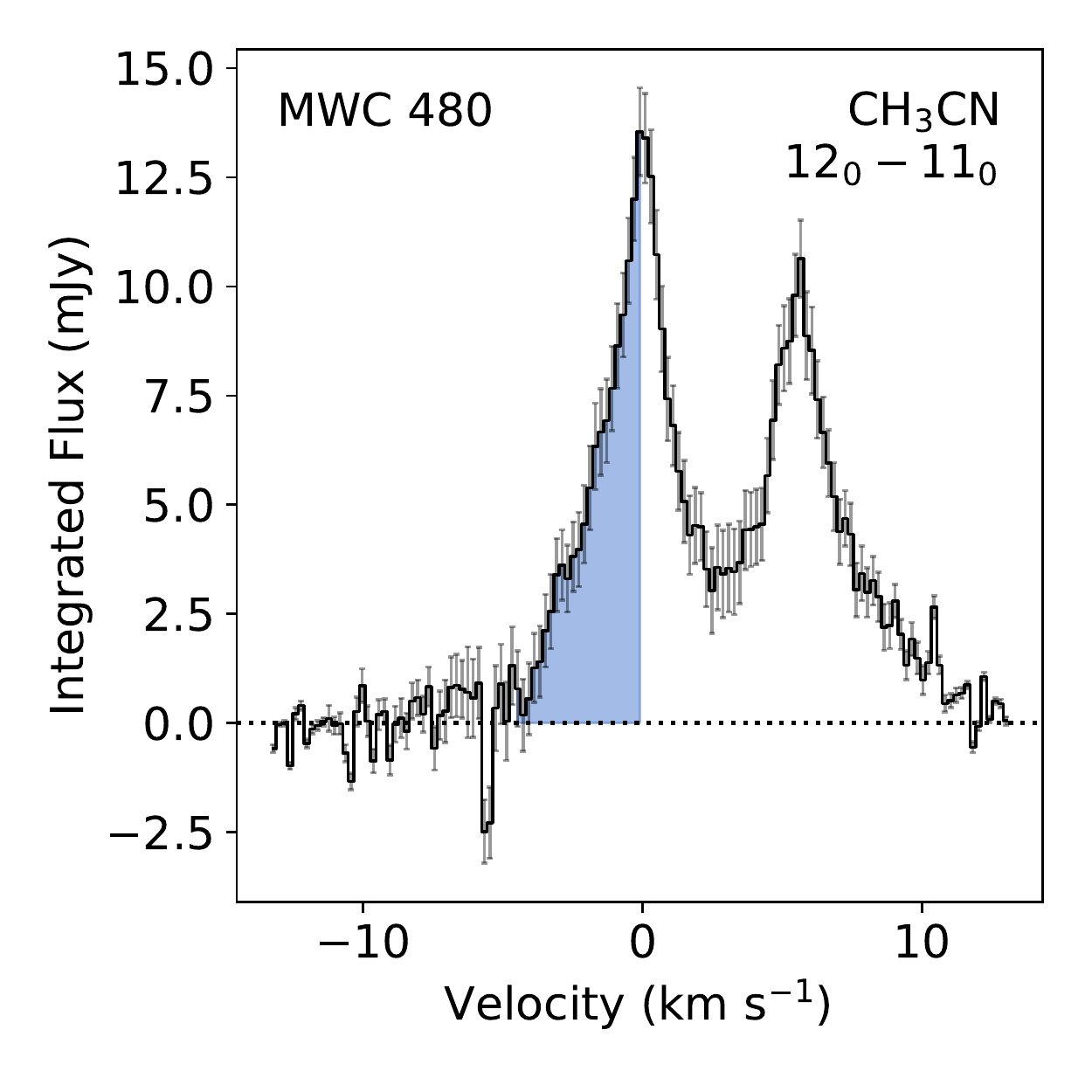}
\includegraphics[width=0.32\textwidth]{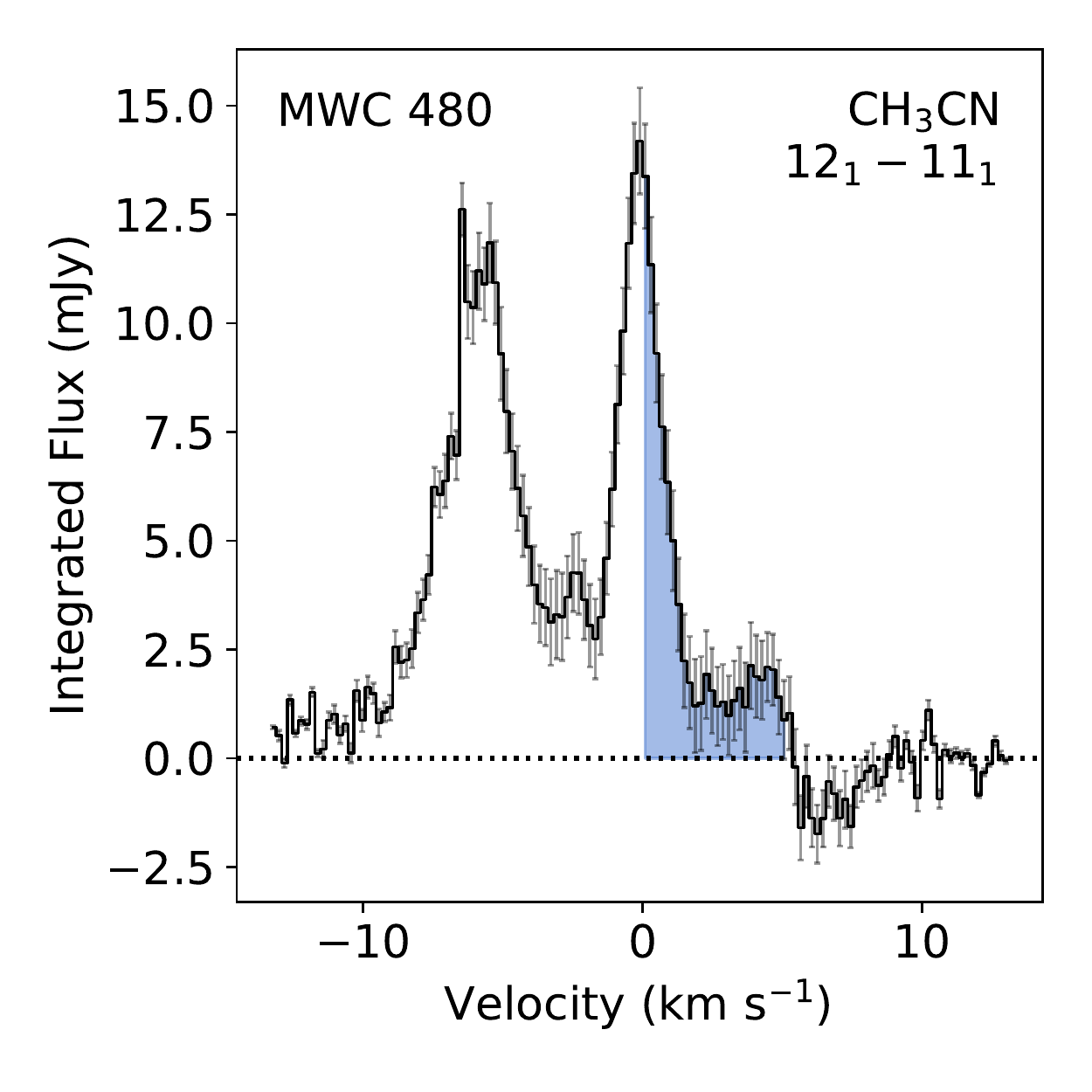}
\caption{Examples of the spectral shifting and stacking procedure for selected lines in the MWC~480 disk in Band 3 (\ce{HC3N} 11--10, left) and Band 6 (\ce{CH3CN} 12--11; middle, right).  {Uncertainties are shown for each channel}, and the shaded region indicates the velocity range across which the line flux was measured (assumed to represent 50\% of the total line flux for the blended \ce{CH3CN} $K=0$ and $K=1$ transitions, see Section \ref{sec:gofish}).}
\label{fig:gofish}
\end{figure*}

\subsection{Imaging}

{Imaging was performed using the \texttt{tclean} task available in \texttt{CASA} version 6 \citep{mcmullin07}.  During cleaning, Keplerian masks were created based on the known geometry of the disks to select regions in position-position-velocity space with emission.  The residual scaling technique of \citet{jorsater95} was applied to the final image cubes to mitigate artefacts introduced by the combination of data from multiple antenna configurations.  Detailed explanations of these steps are provided in \citet{czekala20}.} 

In this work we make use of the fiducial MAPS imaging products.  Given the aforementioned weak nature of many of our targeted lines, coupled with the fact that we compare observations across ALMA Bands 3 and 6, all data presented in this paper utilises images with circularised beams of size $\sim0\farcs3$, obtained by applying a $uv$~taper during imaging \citep[see Section 6.2 of][]{czekala20}.  This was found to provide the optimal combination of angular resolution and sensitivity for our analysis. All image cubes were post-processed in a homogeneous manner through our dedicated pipeline to produce integrated intensity maps (zeroth moment, see Figure \ref{fig:cont_and_best} and Appendix \ref{sec:mom0}) and azimuthally-averaged radial emission profiles \citep[see Section \ref{res:radialprofiles} and][for further details]{law20_rad}.

\subsection{Shifting and stacking} 
\label{sec:gofish} 

In order to robustly recover emission in the image plane from the weakest transitions, we utilise the line shifting and stacking technique available within the \texttt{GoFish} package, making use of the \texttt{integrated\_spectrum()} function \citep{teague_gofish}.  This approach exploits the known geometry and velocity structure of the disk to deproject the rotation profile and combine Doppler shifted emission to a common centroid velocity reference frame.  This results in a single disk-integrated spectrum for each transition that we use to measure a disk-integrated line flux  (or a corresponding upper limit in the case of non-detections).  Velocity ranges for integration are chosen by eye for each transition to include all positive emission in the main core of the line. {Uncertainties are calculated on a per channel basis, taking into account de-correlation along the spectral axis (see also \citealt{yen_2016}).} Figure \ref{fig:gofish} shows examples of the stacking technique for several transitions in MWC~480, and Table \ref{tab:flux} reports these measurements for all transitions and targets. 

We note that the measurement of line flux for the \ce{CH3CN} $K=0$ and $K=1$ transitions is complicated by their separation of only $\sim6$~km\,s$^{-1}$, leading to significant blending at higher velocities.  We attempted to mitigate this by use of a sufficiently tight Keplerian mask but contamination still occurred (as can be seen in Figure \ref{fig:gofish}).  Since these lines are expected to produce symmetric profiles, we therefore adopt the approach shown in Figure \ref{fig:gofish} for these transitions that involves measuring the line flux from the unblended half of the integrated spectrum and assuming symmetry about the rest velocity of the line in order to recover the full line flux.

\subsection{Rotational diagrams}
\label{sec:rotationaldiagram}

We exploit the fact that we have multiple transitions of each molecule to empirically extract both disk-integrated and radial-dependent column densities, $N_\mathrm{T}$, and rotational temperatures, $T_\mathrm{rot}$, wherever possible.  We follow a similar methodology as presented in \citet{Loomis2018a} that we describe here. Under the assumption of optically thin emission, the surface brightness of line emission $I_{\nu}$ is related to the column density of molecules in the upper level of each transition $N_{u}^{\rm thin}$ via
\begin{equation} 
I_{\nu} = \frac{A_{ul} N_{u}^{\rm thin} h c}{4 \pi \Delta v},
\end{equation}   
in which $A_{ul}$ is the Einstein A coefficient and $\Delta v$ is the intrinsic linewidth \citep[see, e.g.,][]{Goldsmith1999}. In the case of disk-averaged emission, $I_{\nu} = S_{\nu}/\Omega$, where $S_{\nu}$ is the flux density and $\Omega$ is the solid angle subtended by the emission, calculated from the radial extent of each line.  Substituting for $I_{\nu}$ and inverting this relation gives  
\begin{equation} N_{u}^{\rm thin} = \frac{4 \pi S_{\nu} \Delta v}{A_{ul} \Omega h c}, 
\end{equation}   
where $S_{\nu} \Delta v$ is the integrated flux density reported for each of the transitions in Table \ref{tab:flux}.

This can then be related to the total column density of the emission through the Boltzmann distribution,
\begin{equation}
    \frac{N_\mathrm{u}}{g_\mathrm{u}} = \frac{N_\mathrm{T}}{Q(T_\mathrm{rot})}e^{-{E_\mathrm{u}/k_\mathrm{B} T_\mathrm{rot}}},
    \label{eqn:nugu}
\end{equation}
where $g_\mathrm{u}$ and $E_\mathrm{u}$ are the degeneracy and energy of the upper level, respectively (see Appendix \ref{sec:molec}), and $Q(T_\mathrm{rot})$ is the partition function at the rotational temperature (linearly interpolated in log-space from calculated values tabulated in the Cologne Database for Molecular Spectroscopy, CDMS; \citealt{cdms}).  Taking the logarithm of Equation \ref{eqn:nugu} yields
\begin{equation} 
\ln \frac{N^{\rm thin}_{u}}{g_u} = \ln N_T - \ln Q(T_{\mathrm{rot}}) - \frac{E_u}{k_\mathrm{B} T_\mathrm{rot}}
\label{eqn:lognugu}
\end{equation}
that can form the basis of a linear least squares regression that derives the rotational temperature, $T_\mathrm{rot}$, and total column density, $N_T$, from the best fitting slope and intercept, respectively. 

In the case that $\tau$ is not negligible, an optical depth correction factor, $C_{\tau}$, must be applied such that the true level populations become
\begin{equation} 
N_u = N_u^{\rm thin} C_{\tau}, 
\end{equation}
where $C_{\tau} = \tau / (1-e^{-\tau})$ and Equation \ref{eqn:lognugu} can then be expressed as
\begin{equation} 
\ln \frac{N_{u}^{\rm thin}}{g_u} + \ln C_{\tau} = \ln N_T - \ln Q(T_{\rm rot}) - \frac{E_u}{k_\mathrm{B} T_{\rm rot}}. 
\label{eqn:lognuguC}
\end{equation} 

The optical depth of individual transitions can be related back to the upper state level populations via
\begin{equation} 
\tau_{ul} = \frac{A_{ul}c^3}{8 \pi \nu^3 \Delta v} N_u (e^{h\nu / k_\mathrm{B} T_\mathrm{\rm rot}} - 1),
\end{equation}   
that implies that $C_{\tau}$ can be written as a function of $N_u$ and substituted into Equation \ref{eqn:lognuguC} in order to construct a likelihood function $\mathcal{L}(data,N_{T}, T_\mathrm{rot})$ to be used for $\chi^{2}$ minimisation.  We use the Markov chain Monte Carlo (MCMC) code \texttt{emcee} \citep{emcee_2013} to fit the observed data using this likelihood function, generating posterior probability distributions that describe the range of possible values of both $N_T$ and $T_\mathrm{rot}$. Best fitting values and their uncertainties are chosen from the median and 16$^{\rm th}$--84$^{\rm th}$ percentile of these posterior distributions, respectively.  We assume uniform priors spanning ranges of $10^{8} < N_T < 10^{15}$\,cm$^{-2}$ and $1 < T_\mathrm{rot} < 150$\,K, respectively, which encompasses the typical values expected for these molecules in disks \citep[see, e.g.,][]{Bergner2018}.

The intrinsic linewidth, $\Delta v$, can be described as a combination of thermal and turbulent broadening such that
\begin{equation}
\Delta v = 2\sqrt{\ln 2} \sqrt{{\frac{2k_{\rm B} T_\mathrm{ex}}{m_{X}}} + t_0^{2} \frac{k_{\rm B} T_\mathrm{ex}}{\mu m_H}},
\label{eqn:linewidth}
\end{equation}
where $m_{X}$ and $m_H$ are the masses of the molecule and hydrogen, respectively, $\mu=2.37$ is the assumed mean molecular weight, and $t_0 \sim 0.01$ is the assumed contribution to the linewidth from turbulence \citep[expected to be small in the line-emitting region of protoplanetary disks, see e.g.,][]{flaherty_2015}.  Since $\Delta v$ is a function of temperature, we iterate the fitting procedure assuming $T_{\mathrm ex} = T_{\mathrm{rot}}$ until the resulting best fitting values for column density and rotational temperature converge, which is usually achieved within $\lesssim 5$ iterations.

For the determination of radially resolved quantities, we perform the same procedure as for the disk-integrated analysis but where $S_{\nu} \Delta v$ is obtained from the radial profiles in annular bins of width $0\farcs075$ (one quarter of a beam) and $\Omega$ is the solid angle of each annulus.  In these cases, the MCMC fitting procedure is performed independently for each bin, and the results combined to report values of $N_T(r)$, $T_\mathrm{rot}(r)$, and $\tau(r)$ for each molecule in each disk.

\begin{figure*}
\includegraphics[width=\textwidth]{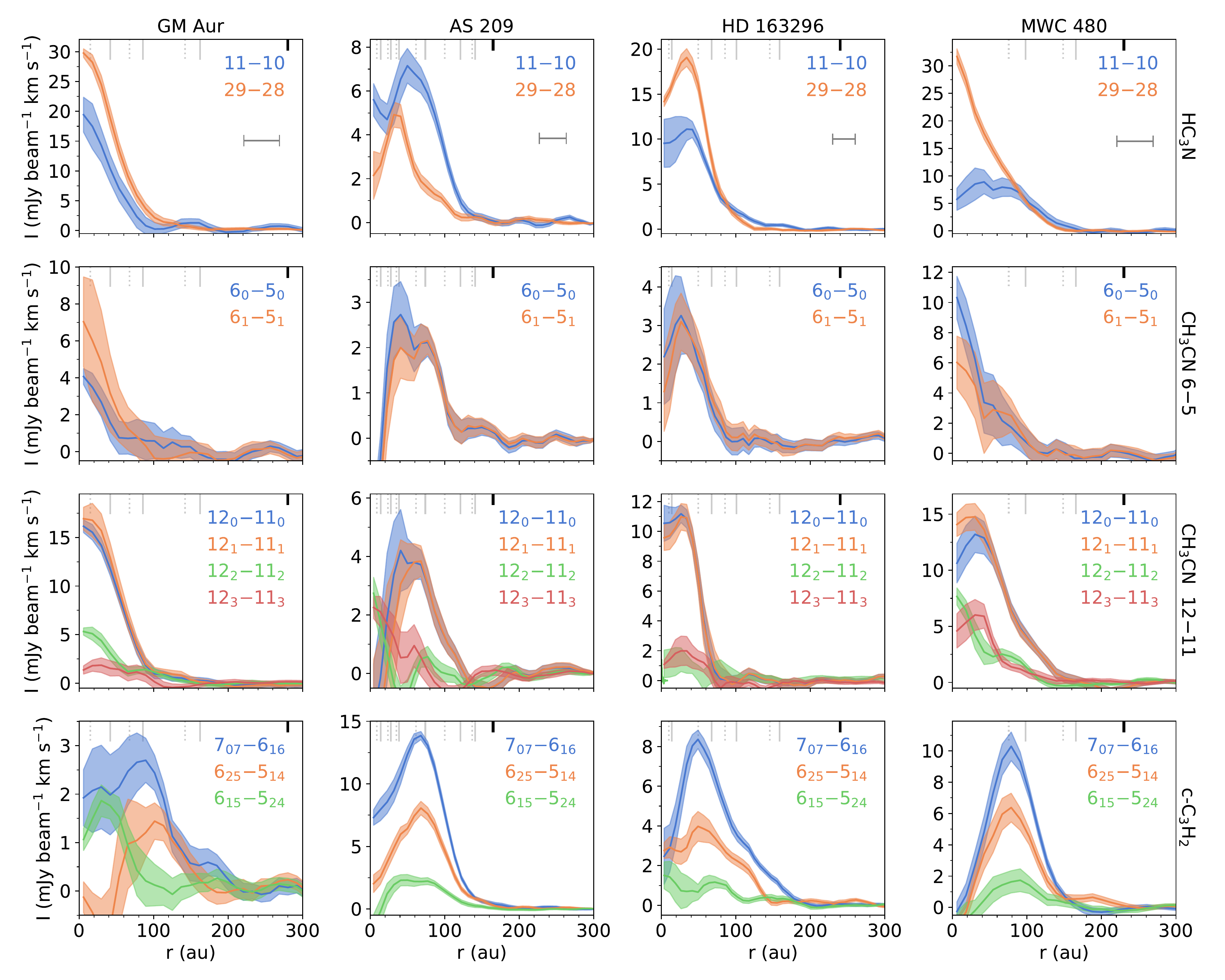}
\caption{Azimuthally-averaged radial emission profiles for \ce{HC3N}, \ce{CH3CN} and $c$-\ce{C3H2} in all sources in which the lines are well detected: AS~209, GM~Aur, HD~163296, and MWC~480.  Locations of continuum rings and gaps are marked with grey vertical lines (solid and dotted, respectively) with the extent of the continuum marked in black (see \citealt{law20_rad}, their Appendix F). {The beam FWHM for each disk is shown in the top panels with a horizontal bar.}}
\label{fig:radial_profiles}
\end{figure*}

\section{Results}
\label{sec:results}

% Include the table... %%%%%%%%%%%%%% 
%%
%%
\input{sigma_flux_table_R1.tex}

%
%%%%%%%%%%%%%%%%%%%%%%%%%%%%%%%%%%%%%

Here we present the results of the matched filter analysis and the imaging, including the generated zeroth moment maps and azimuthally-averaged radial profiles, along with disk-averaged and radially-resolved rotational diagram analysis for each disk. 

\subsection{Matched filter detections}
\label{res:matchedfilter}

As explained in Section~\ref{sec:methods}, all line-containing spectral windows were analysed using a matched filter to confirm the detection of each of our targeted species and lines \citep[see][for full details]{Loomis2018a}.  We used a Keplerian filter that was manually varied in radial extent in to achieve the highest filter response ($\sigma_{f}$) for each molecule.  We found that a 200~au radius filter was optimal for \ce{HC3N} and $c$-\ce{C3H2}, whereas a more compact filter with a radius of 100~au provided the strongest filter response for \ce{CH3CN}, indicating the more compact nature of the emission from this species.  The results of the matched filter are shown in Figure~\ref{fig:filter_all}.

We strongly detect ($\sigma_{f} \gtrsim$ 10) both lines of \ce{HC3N} in four out of five of our sources, but it is not detected in IM~Lup.  We also strongly detect ($\sigma_{f} > 10$) the $7_{07}-6_{16}$ and $6_{25}-5_{14}$ transitions of $c$-\ce{C3H2} in all disks except IM Lup (where there is only a tentative detection of the $7_{07}-6_{16}$ transition).  The $6_{15}-5_{14}$ transition of $c$-\ce{C3H2} is detected with $\sigma_{f} > 5$ in three out of five sources (AS~209, HD163296, and MWC~480).

For \ce{CH3CN} we targeted the $K$-ladder of the $J=6-5$ transition in Band 3 and the $J=12-11$ transition in Band 6. In both cases the $K=0$ and $K=1$ lines are blended. These transitions of the $12-11$ line of \ce{CH3CN} are well detected in all sources except IM Lup for which there is a tentative detection of the $12_1-11_1$ transition only ($\sigma_{f} \sim 3$). Further we detect emission also from the $12_2-11_2$ transition in GM~Aur, HD~163296, and MWC~480.  The emission from \ce{CH3CN} in Band 3 is significantly weaker.  We detect the $6_0-5_0$ and $6_1-5_1$ lines only in AS~209, HD~163296 and MWC~480. There is a tentative detection ($\sigma_{f} \sim 3$) of these transitions as well as the $6_2-5_2$ transition in GM~Aur.

In summary we have detected at least two transitions of each of our targeted species in all disks except IM Lup.  We confirm the previously reported detections of \ce{CH3CN} and \ce{HC3N} in HD~163296 and MWC~480 by \citet{Oberg2015} and \citet{Bergner2018}, and report new detections of these species in GM~Aur and AS~209, as well as a tentative detection of \ce{CH3CN} in IM Lup. Further, we report new detections of $c$-\ce{C3H2} in three sources (GM~Aur, AS~209, and MWC~480) and confirm the previous detection in HD~163296 reported by \citet{Qi2013}.

\subsection{Integrated intensity maps}
\label{res:momentmaps}

Figure~\ref{fig:cont_and_best} presents integrated intensity (zeroth moment) maps for the brightest transitions of \ce{HC3N}, \ce{CH3CN}, and $c$-\ce{C3H2} in our target disks (the full gallery of all transitions in all disks is shown in Appendix \ref{sec:mom0}). 

The $c$-\ce{C3H2} emission has a clear ring-like morphology in AS~209, HD~163296, and MWC~480 for both transitions, potentially indicative of an association with the outer dust rings apparent in the continuum images (see Fig.~\ref{fig:cont_and_best}). 
The zeroth moment maps show that emission from this species is weaker in GM~Aur and only very tentatively present in IM~Lup (confirming the matched filter analysis, see Fig.~\ref{fig:filter_all}).  {\ce{HC3N} also presents a ring-like morphology in both the $J=11-10$ and $29-28$ transitions toward AS~209 and HD~163296, while only the $J=11-10$ transition appears ring-like toward MWC~480 (with the $J=29-28$ appearing centrally-peaked).  GM~Aur initially appears to be an outlier with centrally peaked \ce{HC3N} emission in both transitions, but we note that a ring-like morphology is also observed toward this source when examining higher resolution data \citep[$0\farcs15$, see][]{law20_rad}. In most cases, the $J=11-10$ transition of \ce{HC3N} emission is similarly extended to that of $c$-\ce{C3H2}.  However, the higher energy $J=29-28$ transition ($E_{\rm u} = 190$~K) appears more compact in all disks in which it is well detected.  For \ce{CH3CN}, lines in the $J=6-5$ ladder are significantly weaker than those in the $J=12-11$ ladder.  In contrast to the emission morphology for $c$-\ce{C3H2} and \ce{HC3N}, \ce{CH3CN} appears to have a clear ring-like morphology only in AS~209, with a more centrally-peaked morphology present in the other sources in which it is well detected (although we note the presence of minor dips of emission in the innermost regions of HD~163296 and MWC~480 on scales comparable to the beam).}

\subsection{Radial profiles}
\label{res:radialprofiles}

Radial profiles are particularly powerful at revealing sub-structure not immediately apparent in the integrated intensity maps, and allows us to quantify the radial extent of the emission.  Figure~\ref{fig:radial_profiles} shows the azimuthally-averaged radial profiles of emission \citep[see][for details on how these are generated]{law20_rad} of all lines detected in four of our sources: AS~209, GM~Aur, HD~163296, and MWC~480.  IM Lup possessed numerous non-detections and even those lines that were detected were only done so at tentative significance, so we exclude this disk from the subsequent radial analysis.

Figure~\ref{fig:radial_profiles} confirms both the relatively compact nature of emission from this suite of large organic molecules as well as the ringed morphology present in many sources.  
\ce{HC3N} exhibits either a ringed morphology (in AS~209 and HD~163296) or a centrally compact morphology (GM Aur and MWC~480).  
For AS~209 the lower energy transition ($J=11-10$; $E_{\rm u} = 28.8$~K) is stronger than the higher energy transition ($J=29-28$; $E_{\rm u} = 190$~K), by up to a factor of $\sim$1.5 at their peak positions.
For all other sources, the converse is true, with the higher energy transition between factors of $\sim$1.5--3 stronger at the peak of emission.
In all disks the \ce{HC3N} emission extends out only to the outer edge of the millimetre dust disk.  
For AS~209, the emission from the lower energy transition peaks at $\sim$50\,au that lies between the third and forth millimetre dust rings \citep{Guzmann2018}.
Similarly, both transitions of \ce{HC3N} in HD~163296 peak at $\sim$40\,au that lies between the first and second millimetre dust rings in the high-resolution Band 6 continuum image \citep{Isella2018}.
However, there is a large caveat in this comparison in that the synthesised beam of the high resolution continuum images is significantly smaller than that of the lines (0\farcs03 versus 0\farcs3). 

The radial profiles for \ce{CH3CN} are either centrally peaked (GM~Aur and HD~163296) or display a broad ring (AS~209 and MWC~480) where the latter peaks at radii between $\sim$30--50\,au depending on transition.  In all disks the emission extent is well within the millimetre dust.  In all cases the $12_0 - 11_0$ and $12_1 - 11_1$ transitions are the strongest ($E_{\rm u}$~=~68.9 and 76.0~K, respectively).   {These two Band 6 transitions are considerably stronger than the equivalent Band 3 transitions (i.e., $6_0 - 5_0$ and $6_1 - 5_1$; $E_{\rm u}$~=~18.5 and 25.7~K, respectively) for all disks by factors of 2--3.}  

Finally, the radial profiles for $c$-\ce{C3H2} present a mostly ring-like morphology in all cases. This emission appears to be related to the location of millimetre dust gaps in AS~209, HD~163296 and MWC~480 (at 60, 50, and 80~au, respectively), in agreement with the analysis of the higher resolution ($0\farcs15$) imaging products \citep{law20_rad}.  However, for AS~209 and HD~163296, the $c$-\ce{C3H2} emission peaks outside of that for \ce{HC3N}, highlighting that any correlations between dust continuum and line emission can be different for different molecules. The $c$-\ce{C3H2} emission is confined to within the outer edge of millimetre emission, except for GM~Aur, where there is weak extended emission. In all cases, the $7_{07} - 6_{16}$ transition is the stronger of the two, although this is mainly reflecting the difference in the Einstein A coefficients for the transitions (see Table~\ref{tab:molecular}) as these two transitions have very similar upper energy levels. 
In summary, there is not a one-to-one relation between line emission from a specific species and dust morphology, and nor is there a straightforward correlation between the morphology of emission between different species, except that \ce{HC3N} and \ce{CH3CN} appear broadly similar across each disk. {There are several explanations for the presence of ringed emission in molecular lines in protoplanetary disks \citep[now widely observed, see e.g.][]{Pegues2020, Garufi2021} including a drop in column density in the inner regions due to destructive chemical reactions, an increase in opacity in either dust or lines masking emission from deeper layers in the inner disk regions, or a change in excitation conditions \citep[see, e.g.,][]{vanderMarel_2018, facchini_2018, alarcon_2020}.}  The one constant across the large organic species and disks is that all transitions originate on scales either less than, or comparable to, the extent of the the millimetre dust continuum in each disk.

% Include the table... %%%%%%%%%%%%%% 
%%
%%
\input{disk_int_fits_R1.tex}
%
%%%%%%%%%%%%%%%%%%%%%%%%%%%%%%%%%%%%%

\begin{figure*}[!ht]
\centering
\includegraphics[width=\textwidth,trim=0 0 0 0, clip]{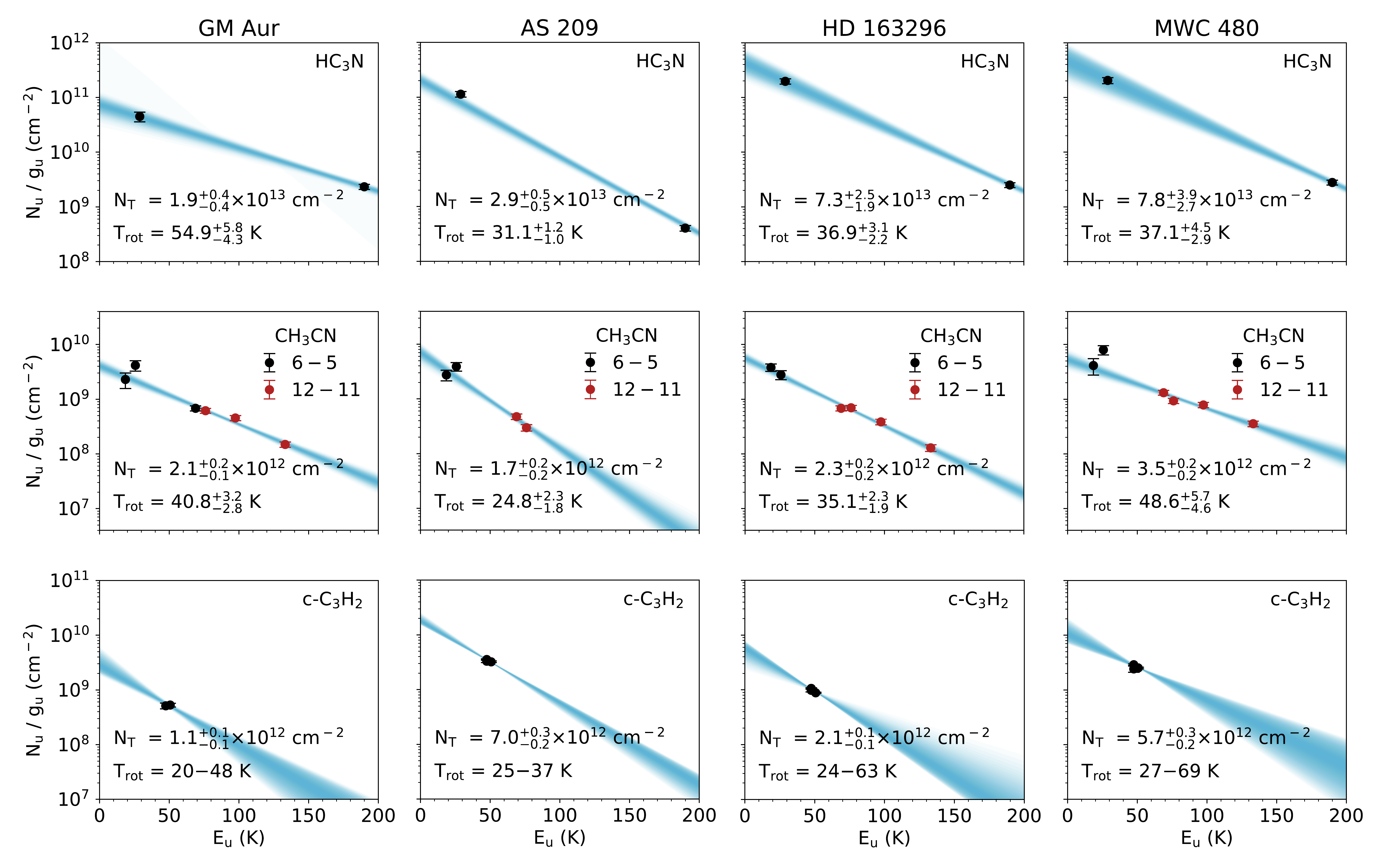}
\caption{Disk-integrated rotational diagrams for \ce{HC3N} (top), \ce{CH3CN} (middle) and $c$-\ce{C3H2} (bottom) labelled with derived $N_{\rm T}$ and $T_{\rm rot}$ values and their uncertainties.  Random draws from the corresponding posterior probability distribution are shown in blue. Due to the limited range of $E_{\rm u}$ spanned by our targeted $c$-\ce{C3H2} transitions, $T_{\rm rot}$ is restricted during fitting as described in Section \ref{sec:diskintfits}.}  
\label{fig:disk_rot}
\end{figure*}

\subsection{Column densities and rotational temperatures}
\label{res:ncolandtrot}

Here we present the results of the calculations of both the disk-integrated and radially-resolved column densities ($N_{T}$) and rotational temperatures ($T_{\rm rot}$) using the methods outlined in Section \ref{sec:rotationaldiagram}. {As noted in Table \ref{tab:flux}, our measurements of the $c$-\ce{C3H2} $7_{07}$--$6_{16}$ line are blended with the corresponding para transition ($7_{17}$--$6_{06}$) at the same frequency \citep[see][]{Spezzano2012}.  For the purposes of the subsequent analysis, we correct the disk-integrated and radially resolved measurements of this line assuming equal contributions from each transition and an ortho-to-para ratio of 3, representative of measurements in other protoplanetary disks \citep[see, e.g.,][]{Guzman2018,TvS2021,Cleeves2021}}.

\subsubsection{Disk-integrated analysis}
\label{sec:diskintfits}

Figure \ref{fig:disk_rot} shows the results of the disk-integrated rotational diagram analysis for the four disks across which multiple transitions of \ce{HC3N}, \ce{CH3CN} and $c$-\ce{C3H2} are detected. For all analysis in which data from multiple observing epochs and bands is used, we assume an additional 10 per cent error to account for uncertainties in flux calibration. All disk-integrated column densities and rotational temperatures are presented in Table~\ref{tab:disk_rot}.  

We note that for $c$-\ce{C3H2}, both targeted transitions are very close in upper energy level. This significantly limits the lever-arm over which to calculate a gradient and thus rotational temperature.  We therefore limit the range of priors for $c$-\ce{C3H2} such that $T_{\rm mid,0} < T_{\rm rot} < T_{\rm atm,0}$, where $T_{\rm mid,0}$ and $T_{\rm atm,0}$ are the midplane and atmospheric temperature at a radial distance of 100~au for each disk, derived from fitting multiple CO isotopologues {\citep[see Appendix \ref{sec:temps}, Table \ref{tab:temps} and][]{law20_vert}}.  

Where detected, the disk-integrated column of \ce{HC3N} ranges from $1.9\times10^{13}$~cm$^{-2}$ for GM~Aur to $7.8\times10^{13}$~cm$^{-2}$ for MWC~480. The disk-integrated rotational temperatures are relatively constant across AS~209, HD~163296 and MWC~480 at $\sim$30--37\,K, but somewhat higher in GM~Aur at 55~K.  It is important to note that the $J=11-10$ transition in all disks appears to be close to optically thick with $\tau = 0.4$--2.8, which may suggest there is a reservoir of \ce{HC3N} in the disks not probed by our observations.

{The disk-integrated column of \ce{CH3CN} presents a narrow range with all values lying within $1.7 - 3.5 \times 10^{12}$~cm$^{-2}$, though a larger range in rotational temperature is seen for the \ce{CH3CN} emission across the disks in the sample with $T_\mathrm{rot} \sim 25 - 49$~K. The calculated optical depths indicate that \ce{CH3CN} emission is optically thin in all disks (though our radially-resolved analysis reveals \ce{CH3CN} may be optically thick across some limited regions of the disks, see Section \ref{sec:diskradfits}).}

{For $c$-\ce{C3H2} there is around a factor of six spread in the disk-integrated column densities ranging from $1.1\times 10^{12}$~cm$^{-2}$ in GM~Aur to $7.0 \times 10^{12}$~cm$^{-2}$ in AS~209.  We again note that the rotational temperatures are constrained based on the temperature structure of the disks rather than population diagrams.  Based on these assumptions, it appears $c$-\ce{C3H2} is optically thin across all disks.}

{The non-detection of \ce{HC3N} and only tentative detections of a single \ce{CH3CN} and $c$-\ce{C3H2} transition toward IM~Lup prevents any determination of well constrained values for rotational temperature or column density for these molecules.  However, by performing the disk-integrated analysis where non-detections provide upper limits, we are able to determine the corresponding upper limit to column density for these molecules in the IM~Lup disk (see Table \ref{tab:disk_rot}).  The resulting \ce{HC3N} column density of $\lesssim5.5\times10^{12}$~cm$^{-2}$ is a factor of 3--15 times lower than values found in the other disks.  Similarly, the upper limit for the column of \ce{CH3CN} and $c$-\ce{C3H2} in IM~Lup are factors of 3--5 and 2--8 times lower than for the other disks, respectively.}

\subsubsection{Radially-resolved analysis}
\label{sec:diskradfits}

\begin{figure*}[htp]
\centering
\includegraphics[height=0.33\textwidth,trim=0 0 0 0, clip]{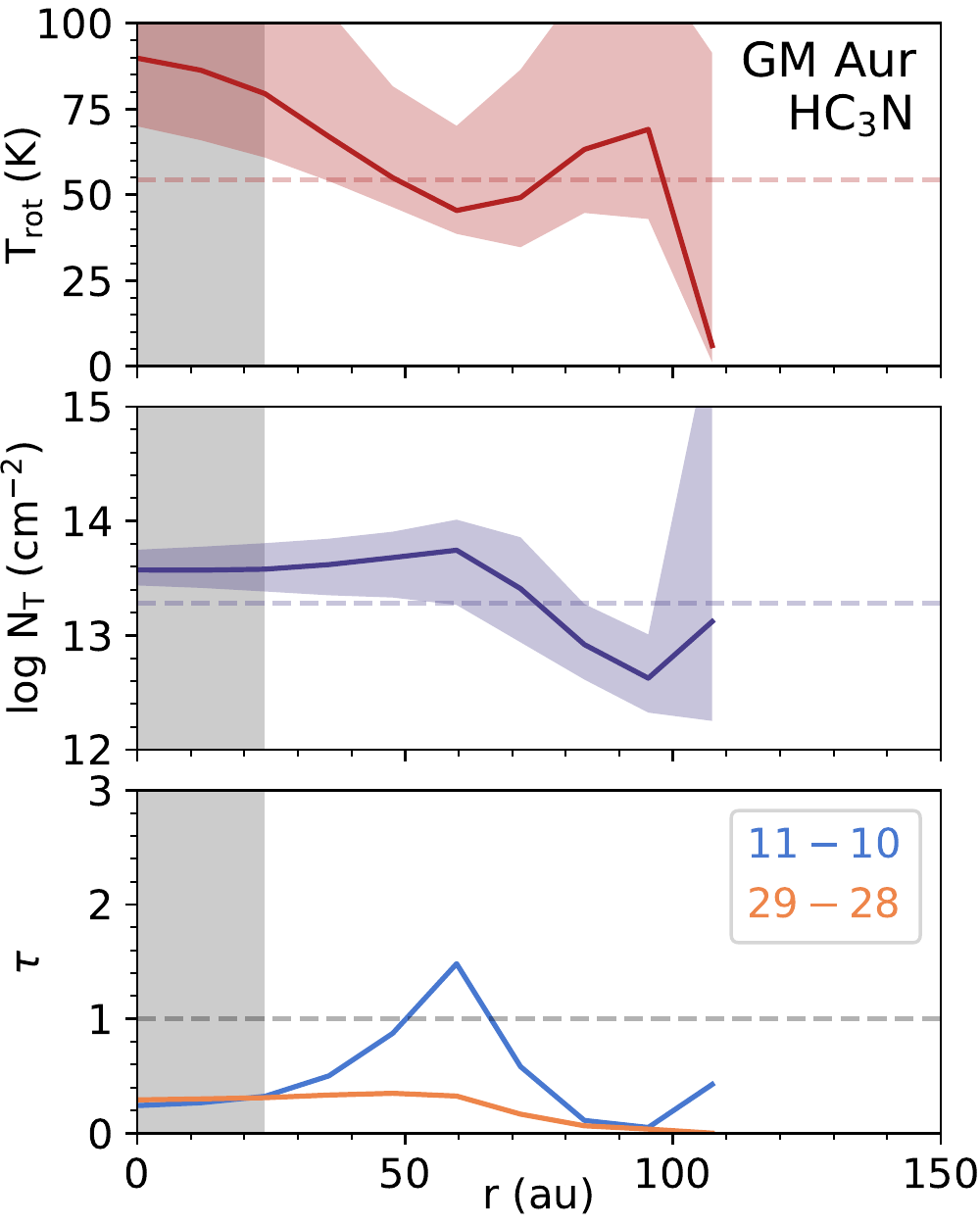}
\includegraphics[height=0.33\textwidth,trim=1.24cm 0 0 0, clip]{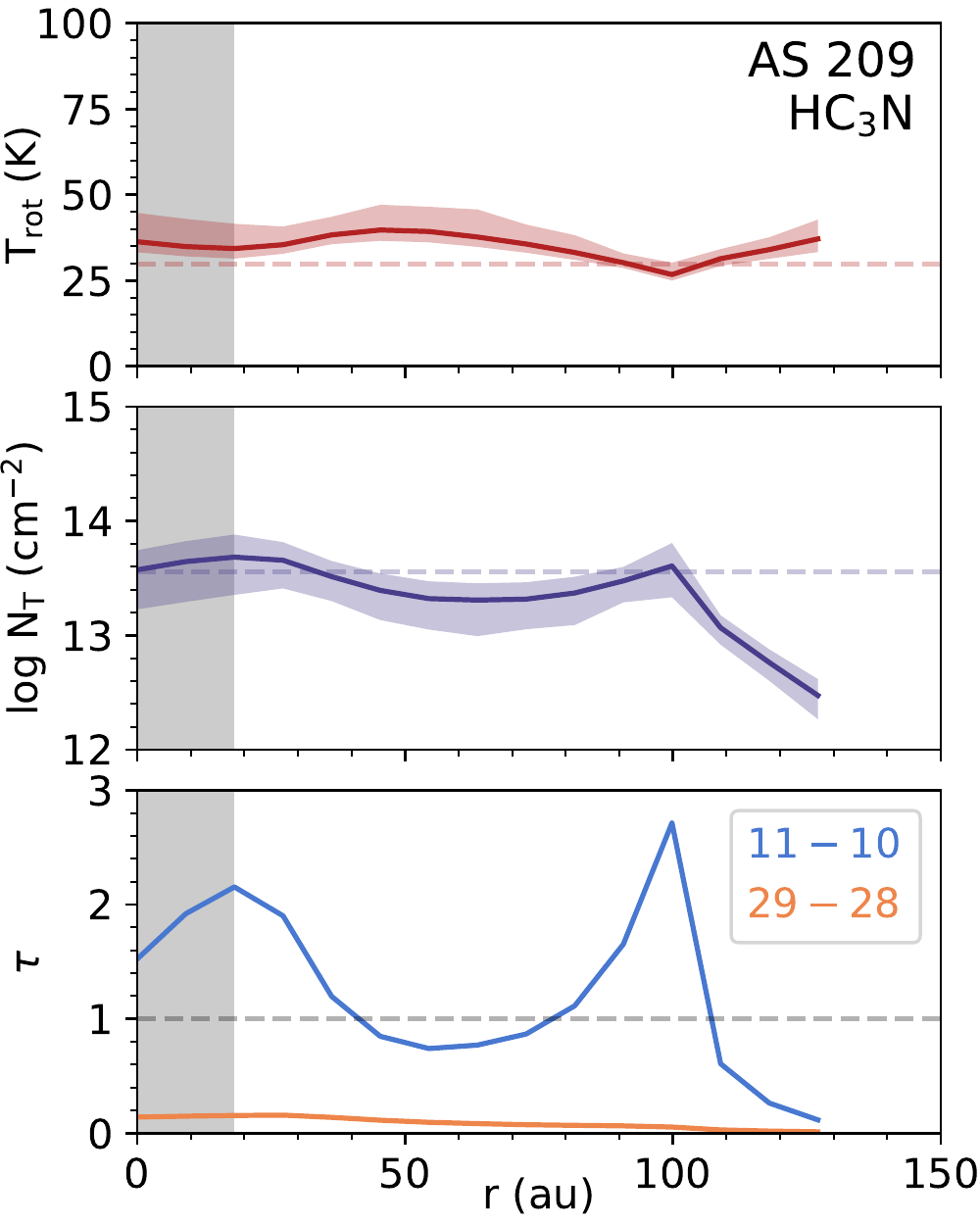}
\includegraphics[height=0.33\textwidth,trim=1.24cm 0 0 0, clip]{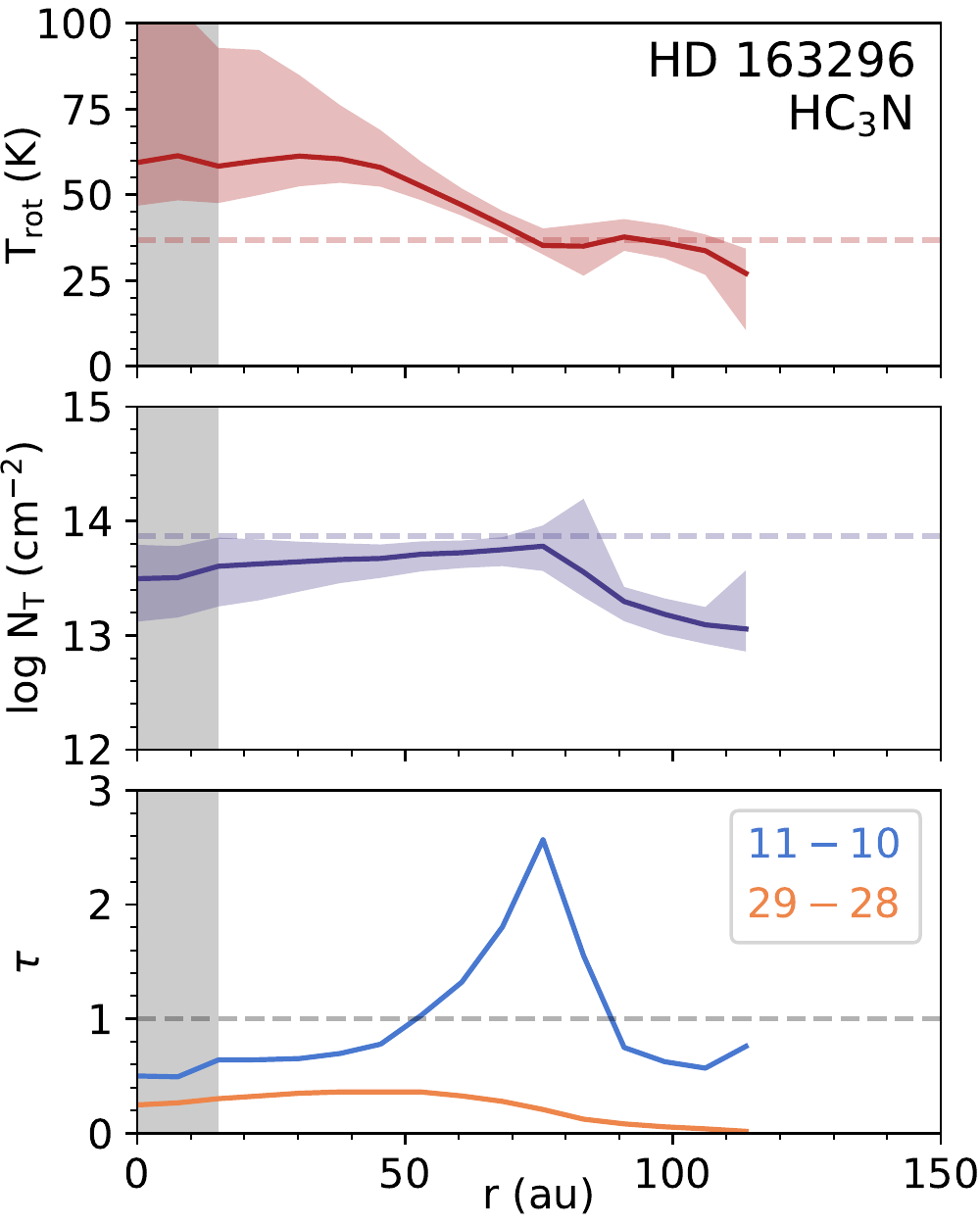}
\includegraphics[height=0.33\textwidth,trim=1.24cm 0 0 0, clip]{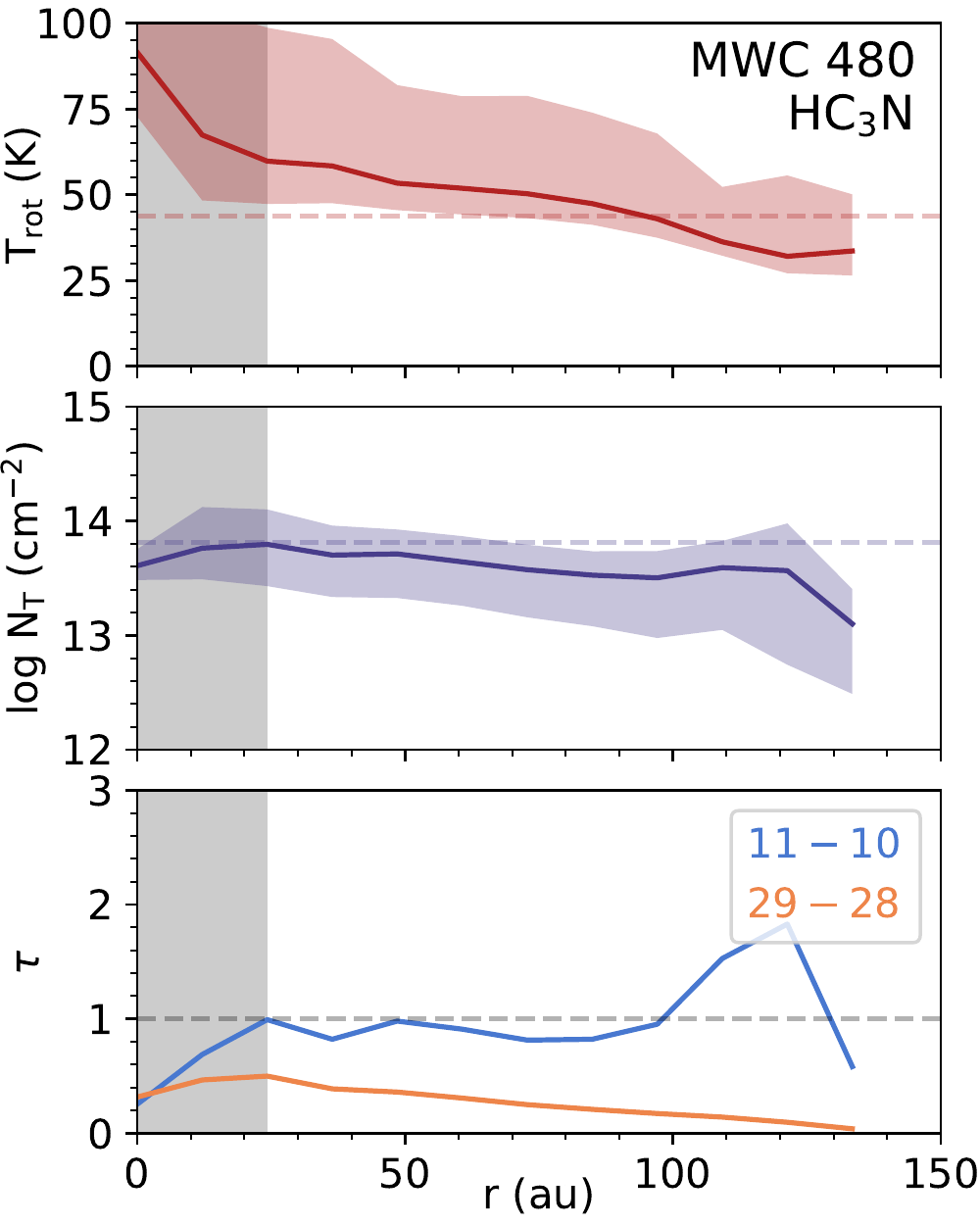}

\vspace{0.5cm}

\includegraphics[height=0.33\textwidth,trim=0 0 0 0, clip]{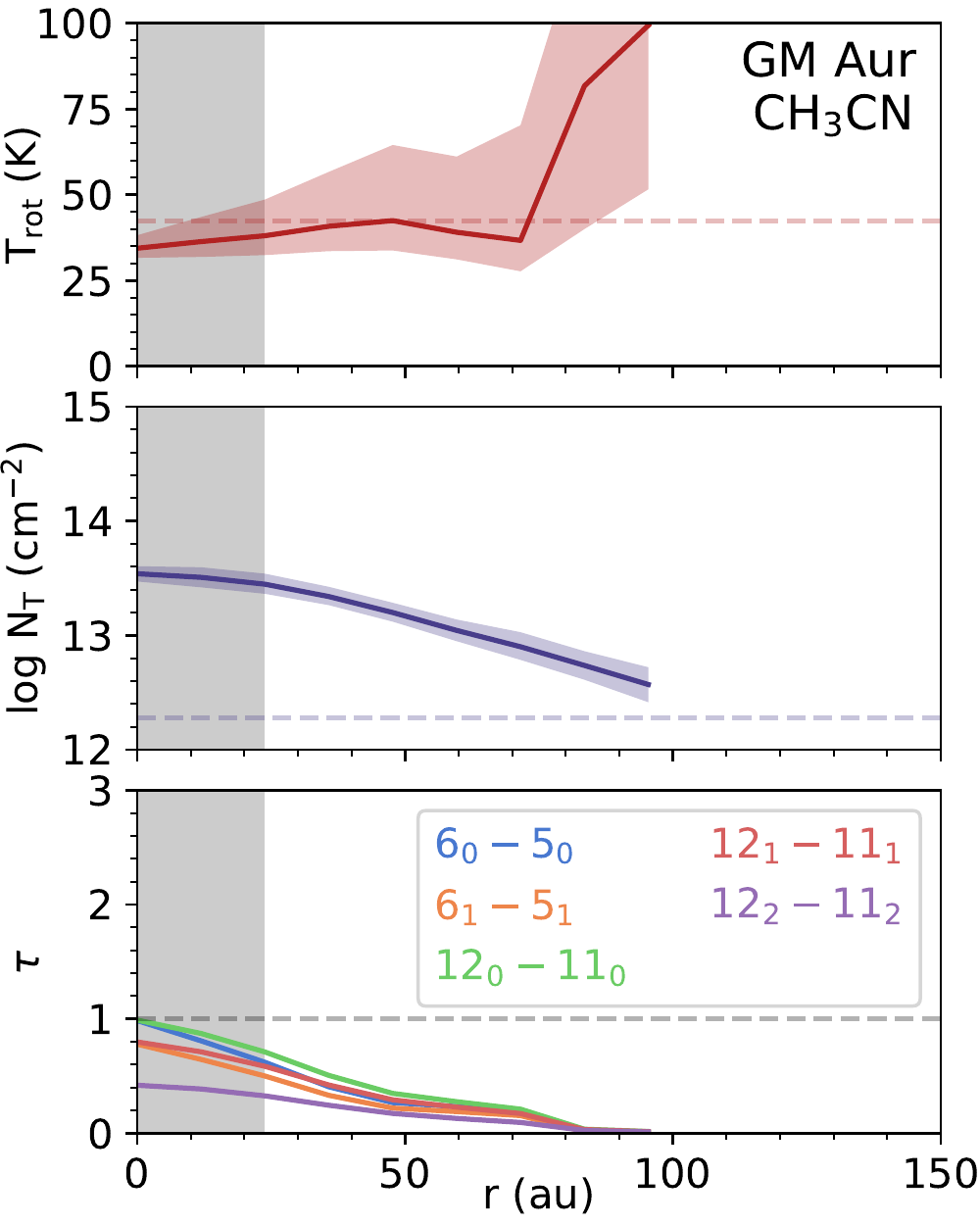}
\includegraphics[height=0.33\textwidth,trim=1.24cm 0 0 0, clip]{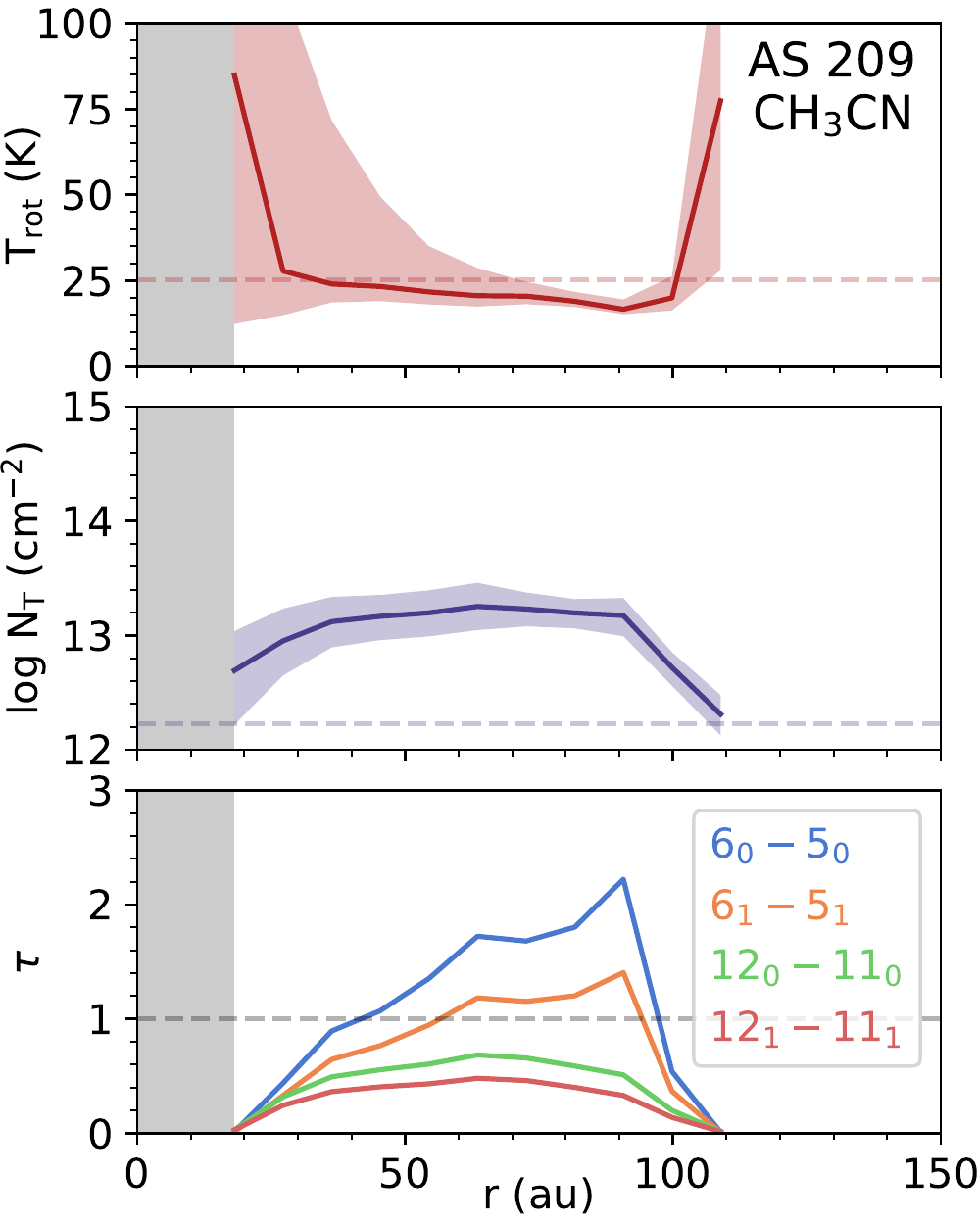}
\includegraphics[height=0.33\textwidth,trim=1.24cm 0 0 0, clip]{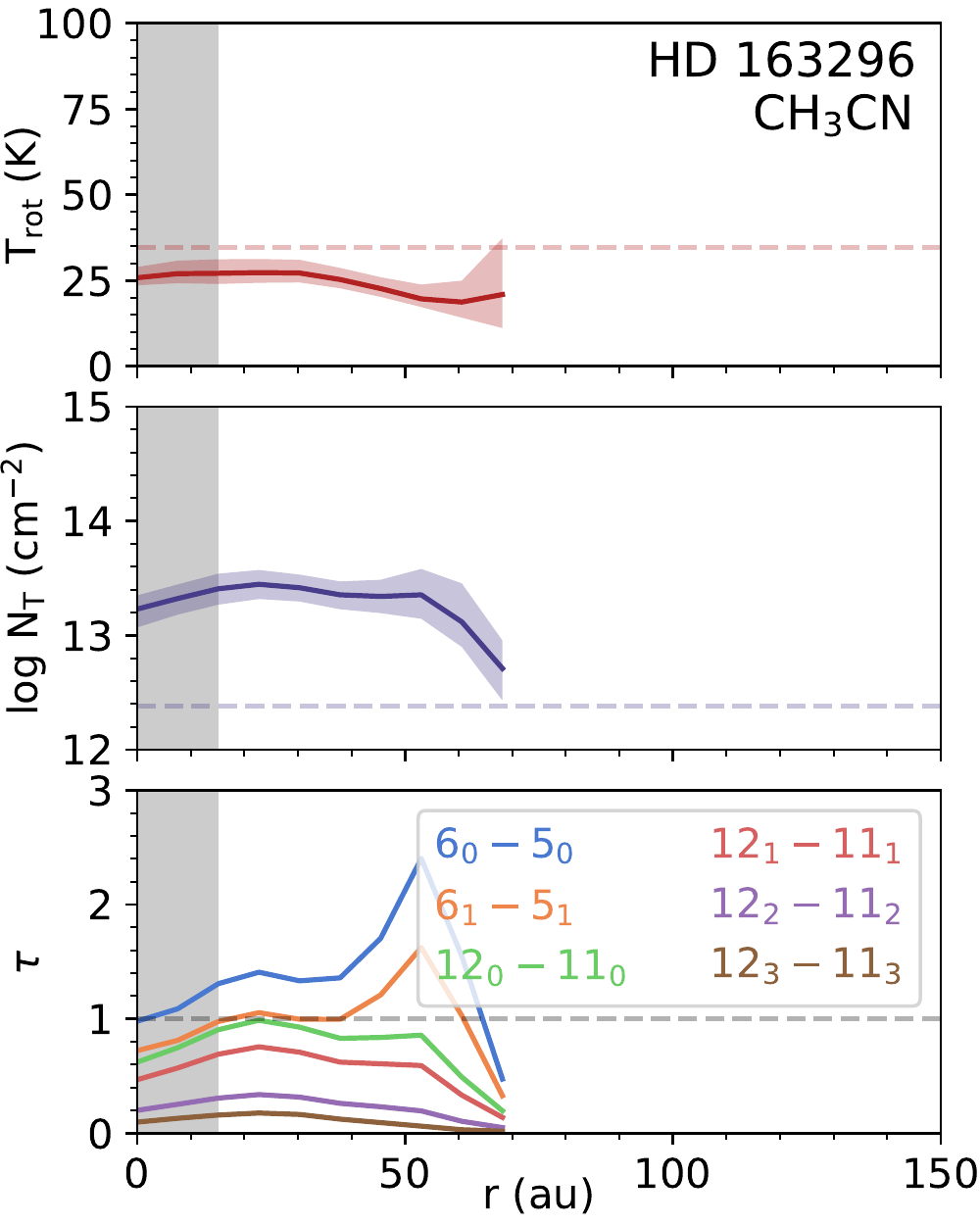}
\includegraphics[height=0.33\textwidth,trim=1.24cm 0 0 0, clip]{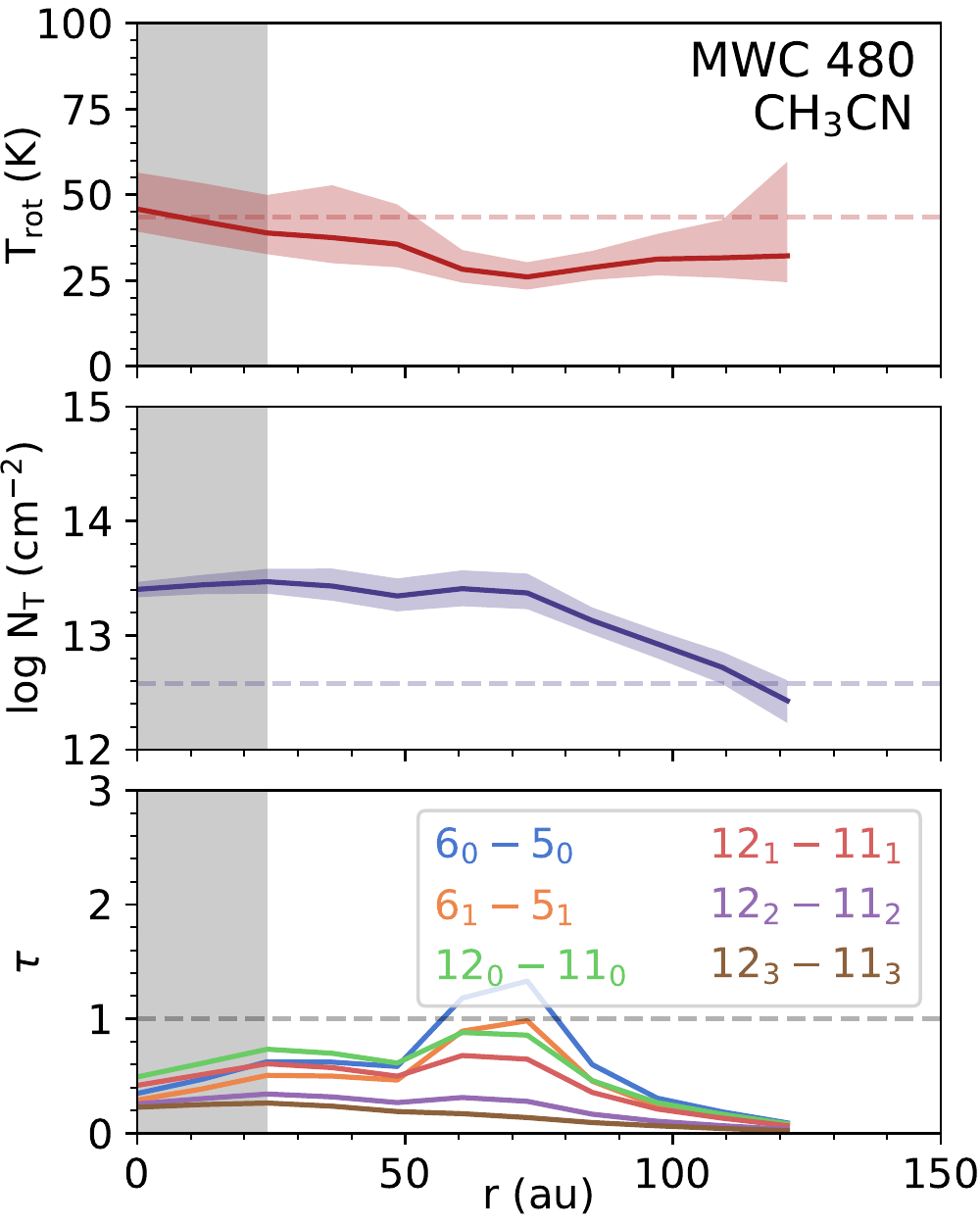}

\vspace{0.5cm}

\includegraphics[height=0.33\textwidth,trim=0 0 0 0, clip]{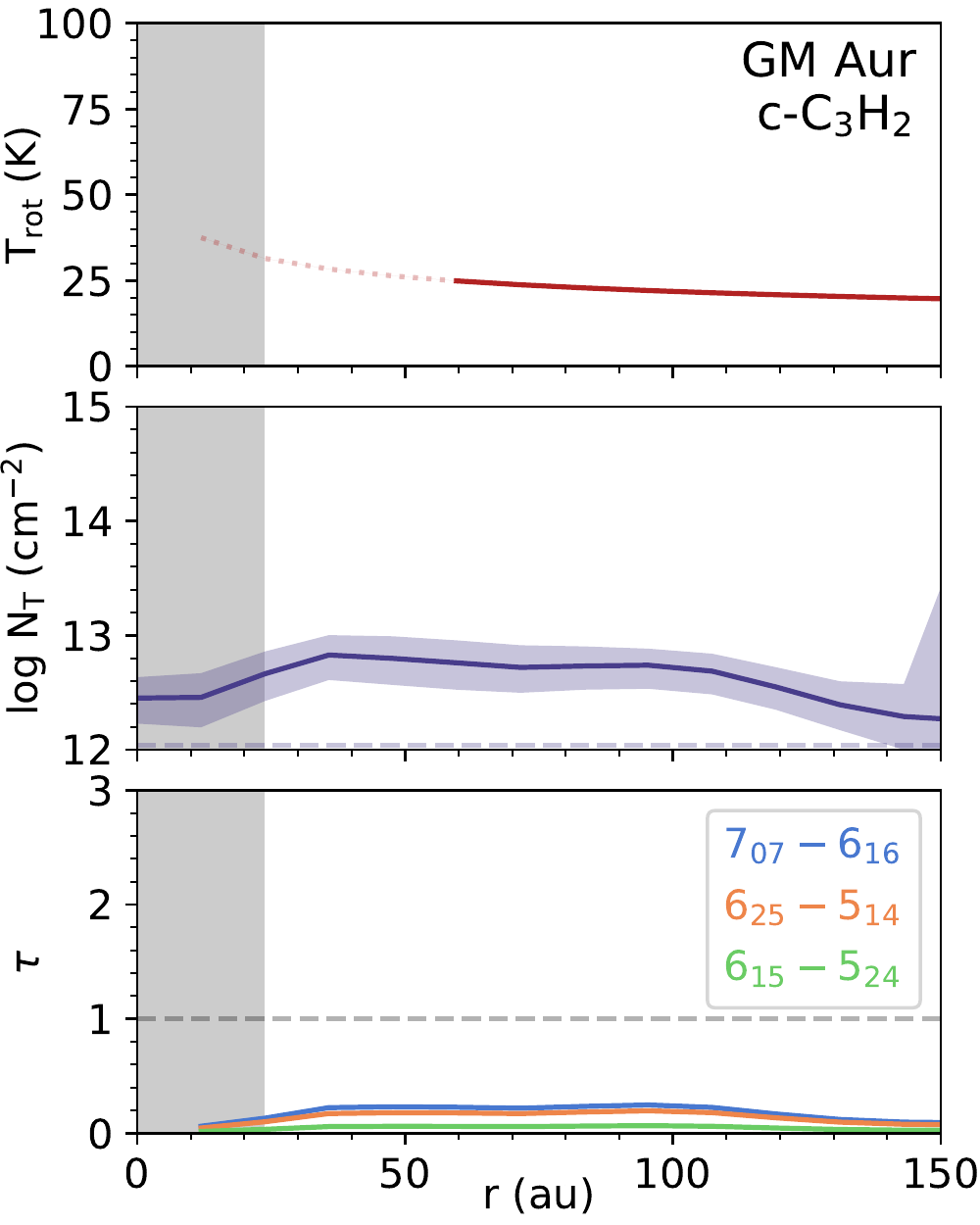}
\includegraphics[height=0.33\textwidth,trim=1.24cm 0 0 0, clip]{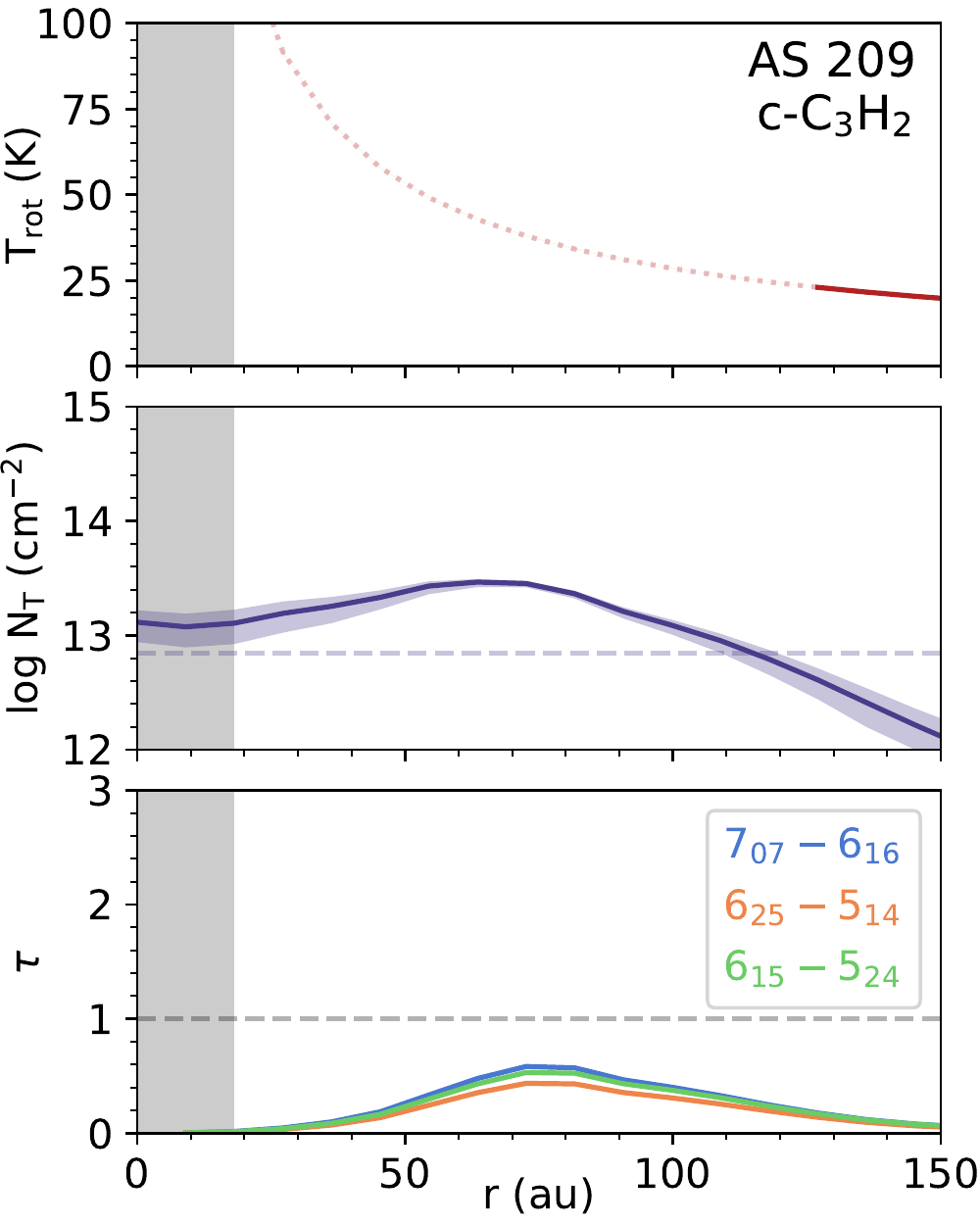}
\includegraphics[height=0.33\textwidth,trim=1.24cm 0 0 0, clip]{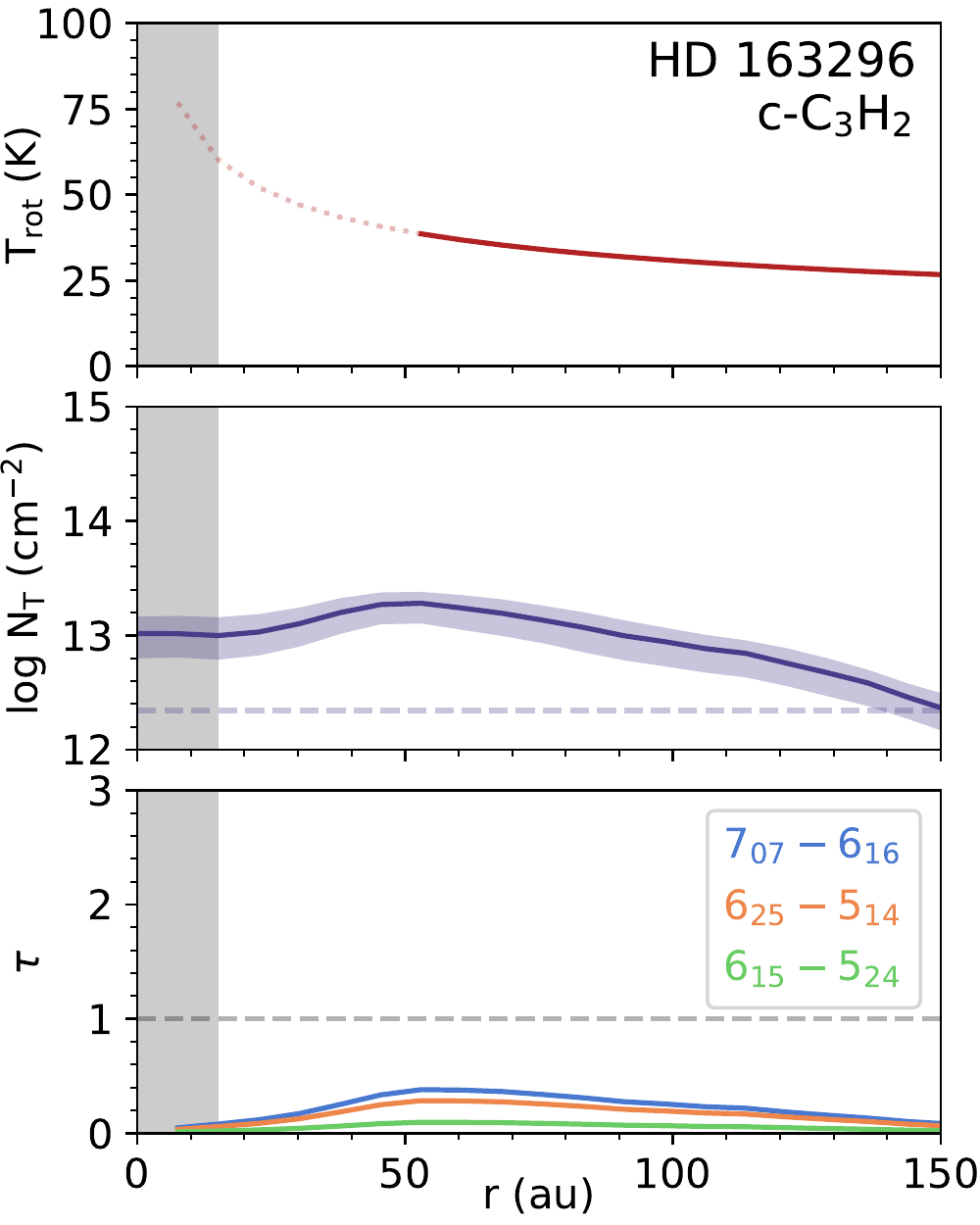}
\includegraphics[height=0.33\textwidth,trim=1.24cm 0 0 0, clip]{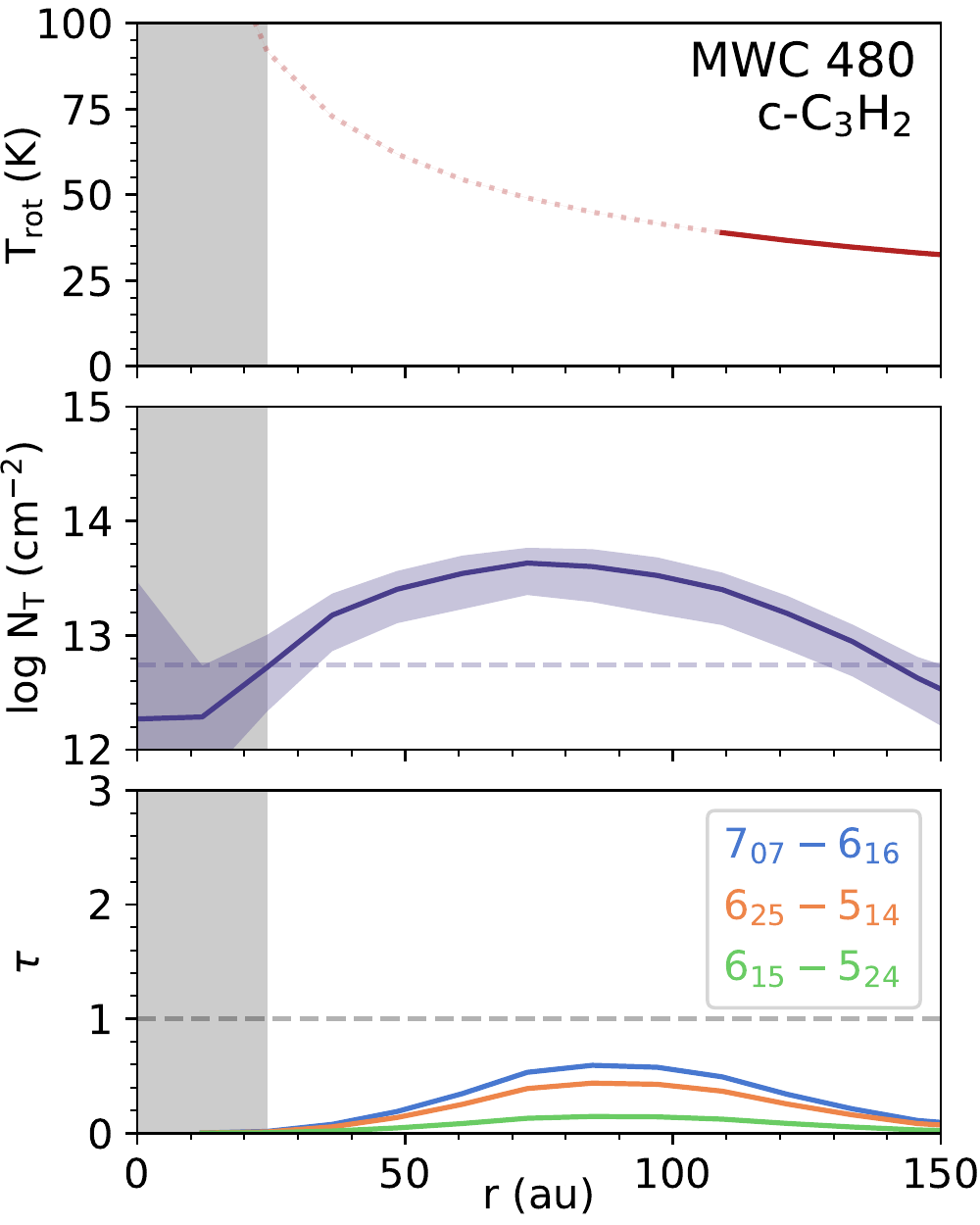}

\caption{Radially-resolved rotational diagrams for \ce{HC3N} (top), \ce{CH3CN} (middle) and $c$-\ce{C3H2} (bottom).  Dashed lines indicate median values derived from the disk-integrated analysis (see Table \ref{tab:disk_rot}) and $\tau=1$.  Grey shaded regions indicate a radial extent of $0\farcs15$ (half of the beam size) within which quantities should be treated with caution. For $c$-\ce{C3H2}, $T_{\rm rot}$ is fixed to the power-law fit of the \ce{^{13}CO} $J=2-1$ temperature from \citet{law20_vert}, with extrapolated values shown with a dotted line (see Section \ref{sec:diskradfits}).}  
\label{fig:radial_rot}
\end{figure*}

The disk-integrated analysis demonstrates interesting similarities and differences across the molecules and sources in our small sample. Since in most cases the emission is spatially resolved, this allows us to also conduct a radially resolved excitation analysis to examine radially-dependent variations that cannot be probed by a disk integrated analysis. Figure \ref{fig:radial_rot} shows the results of this analysis presenting the radial-dependent rotational temperature, column density, and optical depth profiles for all molecules and sources in which emission is well detected. 

For \ce{HC3N}, the radial column densities reveal some structure, with a plateau of higher column densities typically reached in the inner ($\lesssim$100~au) region ($\sim 10^{14}$~cm$^{-2}$) declining monotonically with radius to $\sim 10^{13}$--$10^{12}$~cm$^{-2}$ beyond $\sim$50--100~au.  
The rotational temperature is also relatively constant across all disks at $\sim$50~K.  
There is an indication of an increase in rotational temperature within the inner 20--30 au of GM~Aur and MWC~480, although derived quantities in these regions may be affected by issues such as beam smearing.  In addition, the emission from the $J=11-10$ transition of \ce{HC3N} appears to be optically thick throughout a significant radial region of each disk, sometimes as far out as 130~au. 
Hence, it is possible that there is a significant reservoir of \ce{HC3N} in these regions of these disks to which our observations are not sensitive, and so our radially-resolved column densities are likely lower limits in these regions.

For \ce{CH3CN} there are more differences seen between disks in the radially resolved column densities than for \ce{HC3N}.  
This is despite the disk-integrated analysis suggesting very similar disk-integrated columns across all four sources.  
In general, the radially resolved column densities are between a factor of 5--10 higher than those derived from the disk-integrated analysis.
For GM~Aur and MWC~480 the column density monotonically decreases with radius from $\sim$5$\times10^{13}$~cm$^{-2}$ to a few $10^{12}$~cm$^{-2}$ at approximately 100~au. 
On the other hand, the column density of \ce{CH3CN} in AS~209 and HD~163296 has more of a broad ring-like structure.  
The rotational temperatures of \ce{CH3CN} appear to show a similar radial behaviour across all sources, remaining relatively constant at 30--40~K within $\sim$100\,au, which is similar to values derived from the disk-integrated analysis.  There are hints of a rise to $\sim$80~K in the outer regions of GM~Aur and AS~209, but this may be due to a limited signal-to-noise ratio in these regions. {There is also a significant rise in rotational temperature in the inner region of AS~209 to $\sim$80~K, as expected based on the weak $K=0$ and $K=1$ lines across this region (although we note that the uncertainties here are large).} Our analysis also reveals that the \ce{CH3CN} Band 3 lines, in particular the $6_0-5_0$ and $6_1 - 5_1$ transitions, approach optical depths of 1 or higher out to 100~au in three of the disks (AS~209, HD~163296 and MWC~480) and in the inner regions ($\lesssim$20~au) of GM~Aur.  This indicates that, similarly to \ce{HC3N}, our observations may not be sensitive to the bulk of the \ce{CH3CN} emitting material.

For the radially resolved analysis of $c$-\ce{C3H2}, we are again limited in our calculation of a rotational temperature by the small difference in upper energy level of the targeted transitions.  We therefore fix the rotational temperature to the one-dimensional power law profile, $T(r) = T_{100} \times \left(\frac{r}{\rm{100\,au}} \right)^{-q}$, fitted to the \ce{^{13}CO} $J=2-1$ brightness temperature for each disk (see Table \ref{tab:temps} in Appendix \ref{sec:temps} and \citealt{law20_vert}), that we extrapolate to all radii covered by our $c$-\ce{C3H2} emission. 
In all cases the emission is optically thin, or moderately optically thick ($\tau \sim 0.6$ between 70--100~au in AS~209 and MWC~480).  
The column density for GM~Aur is broadly constant at $\sim 10^{13}$~cm$^{2}$, while AS~209, HD~163296 and MWC~480 show broad peaks at values of $\sim$5--$8\times10^{13}$~cm$^{-2}$ and monotonic decreases with radius beyond $\sim$100~au.

In summary, our radially resolved analysis shows much more variation across the sources than suggested by a simple disk integrated analysis.  Of particular note are the larger peak column densities extracted for \ce{CH3CN} and $c$-\ce{C3H2}, and the high optical depth of the 11--10 and 6--5 transitions of \ce{HC3N} and \ce{CH3CN} respectively, indicating that for these molecules the derived column densities in the inner disk may also be lower limits.

\section{Discussion}
\label{sec:discussion}

The large range of information available from the multiple species and transitions targeted with MAPS enables us to probe further into the origin of the large organic molecules studied here.

\subsection{Distribution \& abundance of the large organics}
\label{sec:origin}

\begin{figure*}[!ht]
\centering
\includegraphics[width=0.49\textwidth]{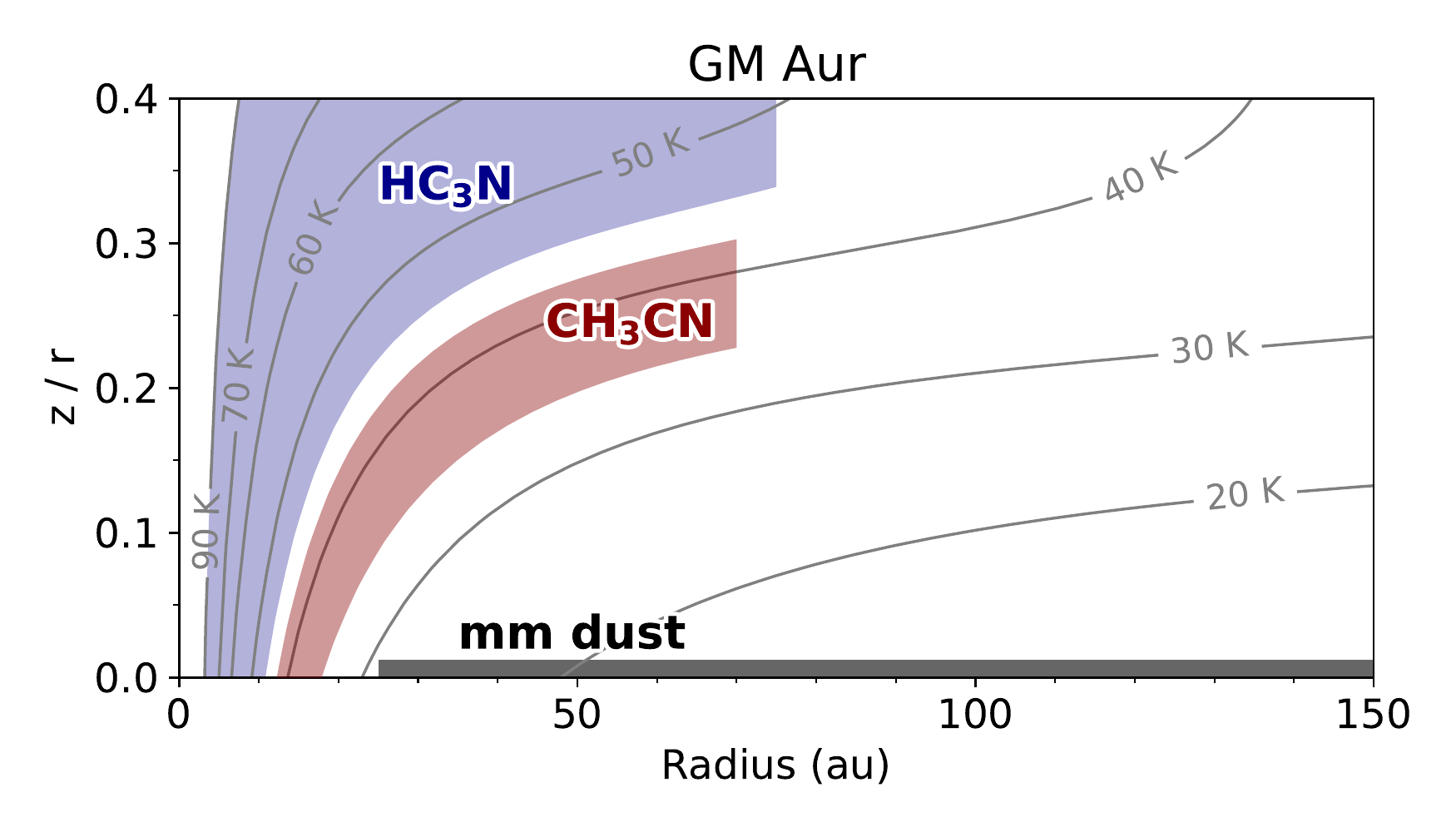}
\includegraphics[width=0.49\textwidth]{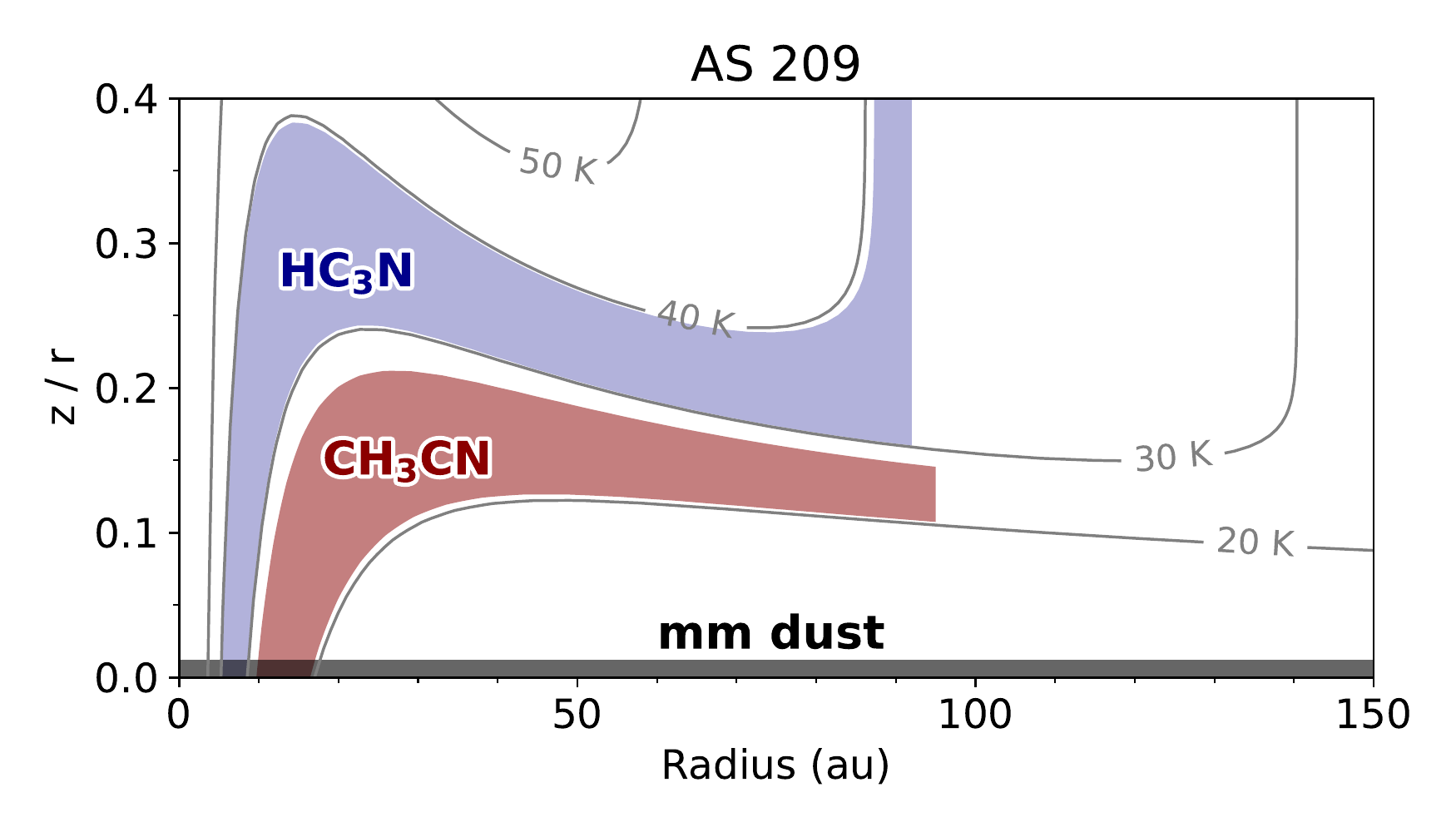}
\includegraphics[width=0.49\textwidth]{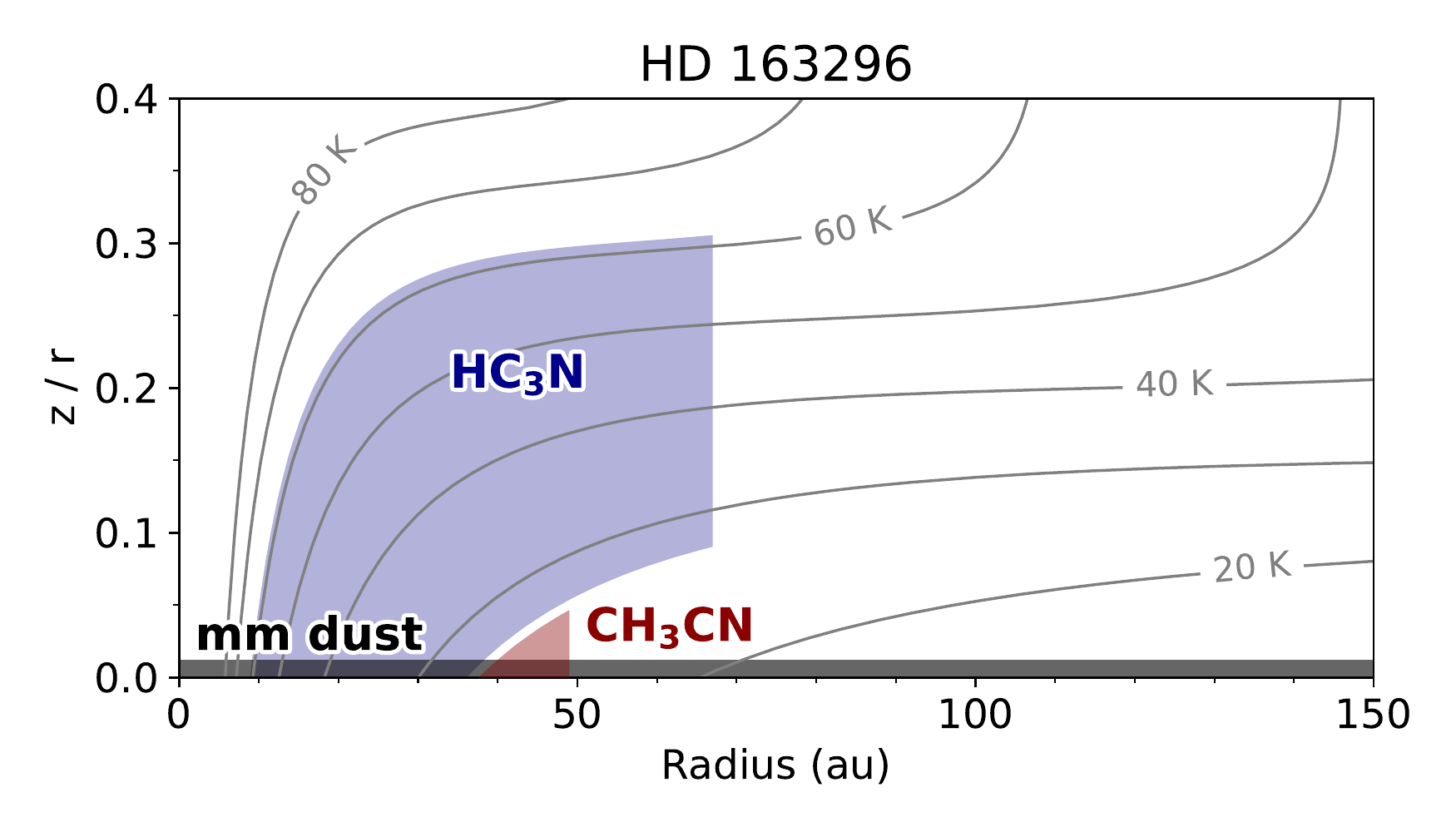}
\includegraphics[width=0.49\textwidth]{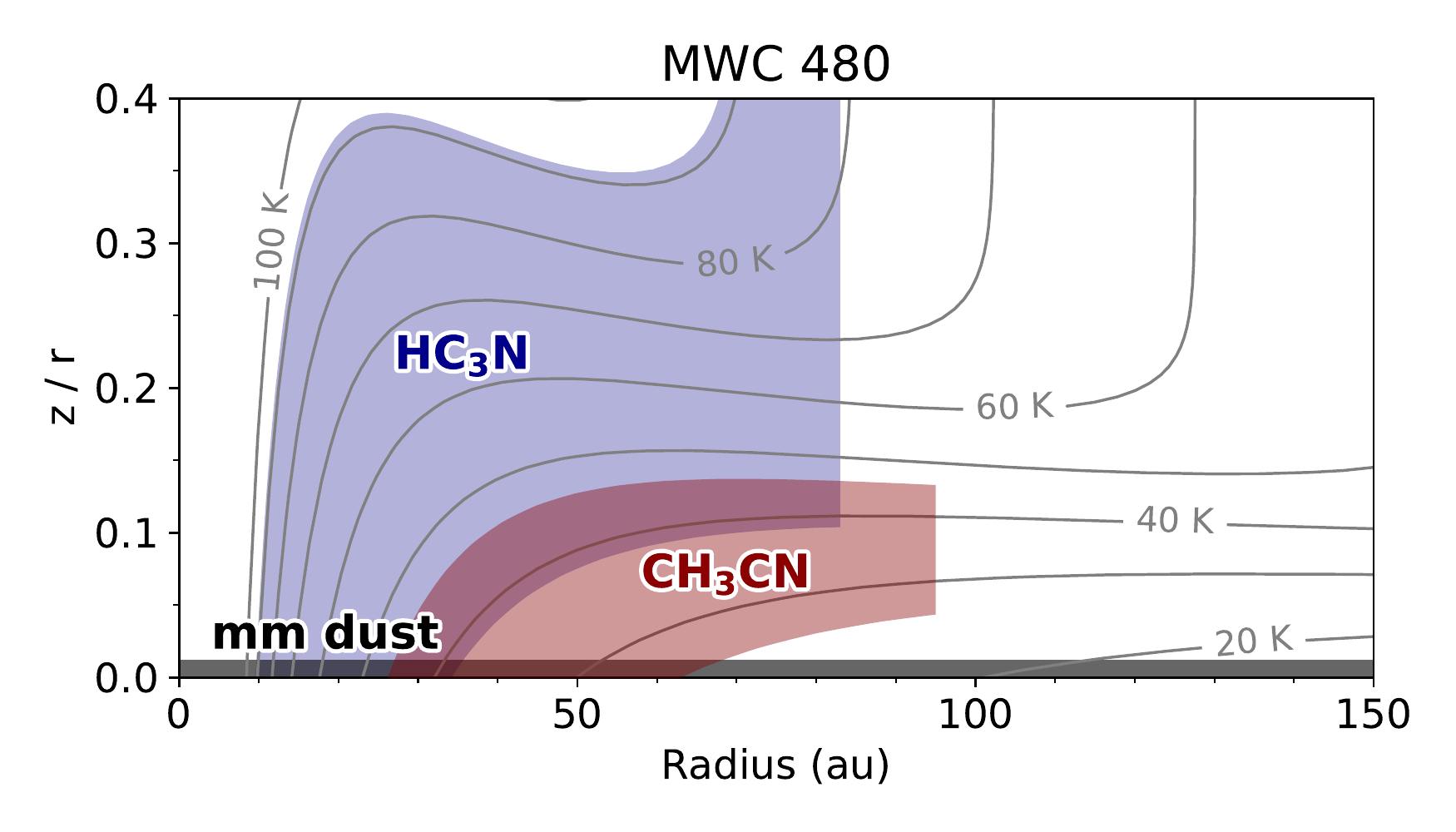}
\caption{Two-dimensional disk temperature structures (contours) overlaid with the region of emission for \ce{HC3N} and \ce{CH3CN} along with the radial extent of the millimetre dust continuum emission at 220~GHz (see Section \ref{sec:origin}).}
\label{fig:origin}
\end{figure*}

{We can use the results of our radially-resolved rotation diagram analysis to further examine the origin of the emission from the large organic molecules.  \citet{law20_vert} have determined two-dimensional temperature structures, $T(r,z)$, for each of our target disks based on fitting brightness temperatures from the mostly optically thick $J=2-1$ transitions of \ce{^{12}CO}, \ce{^{13}CO} and \ce{C^{18}O} isotopologues, based on the two-layer model of \citet{Dullemond2020}.  We can use this temperature structure to examine the  spatial origin of the emission from these molecules in each of our disks.  Under the assumption that the rotational temperature of each molecule is a measure of its true temperature (a sensible approximation given the high densities in protoplanetary disks) we can use the range of derived $T_{\rm rot}$ for each molecule from Figure \ref{fig:radial_rot} to map the range of heights in the disk from which the emission would originate.  We can determine the radial range of emission following the approach of \citet{law20_rad} by calculating the radius encompassing 90\% of flux in the radial profile from each molecule.  Here we measure this using the $0\farcs$3 data for consistency with our previous analysis, but these only show minor differences with the results of \citet{law20_rad}.  Combining these radial ($r$) and height ($z$) bounds, we can determine the two-dimensional origin of the emission for each molecule, which is shown for each disk in Figure \ref{fig:origin}.} 

We note that the original $T(r,z)$ structure derived in \citet{law20_vert} is not a sensitive probe of the midplane temperature structure of the disks.  This is because the CO emission from which these temperature structures are calculated originates from intermediate-to-high relative heights in the disks ($z/r \gtrsim 0.1$), and temperatures below 20\,K were excluded from the fitting procedure. We have therefore re-normalised the $T(r,z)$ structures such that the CO snowline (assumed to be 20\,K) is reached at a midplane radius consistent with \citet{zhang20}, whose thermochemical models reproduce both the SED and radial CO column density of each disk.  We also note that we exclude the $c$-\ce{C3H2} emission from this comparison, since the rotational temperature we assume is not independent of the temperature structure derived from CO in each disk.

{Our analysis of the origin of the line emission in $T(r,z)$ space reveals common trends across the four disks.  In all cases, the \ce{HC3N} emission appears to originate from the highest temperature regime of the two molecules ($\sim 30$--80\,K), and therefore at higher relative heights, generally in the range of $z/r=0.1$--0.4.   In contrast, the lower measured temperature of \ce{CH3CN} means it originates from lower relative heights, generally $z/r \lesssim 0.1$--0.2.  The underlying differences in temperature structure between the warmer disks around Herbig stars (HD~163296 and MWC~480) and the cooler disks around T Tauri stars (GM~Aur and AS~209) result in differences in the location of the emitting region of the molecules.  In the cooler disks, the large organics primarily trace the warm molecular layer, and may only trace the disk midplane interior to 20~au.  In the warmer disks, the large organics trace a wider radial range at the midplane interior to 60~au, and the warm molecular layer exterior to this.  In particular, \ce{CH3CN} appears to originate almost exclusively from $z/r < 0.1$ in these warm disks.  Nevertheless, the general morphology of the region of emission is similar for all disks -- a region close to the midplane interior to the \ce{CO} snowline, and an elevated warm molecular layer beyond it.  Figure \ref{sec:origin} also shows the extent of the 220~GHz dust continuum emission for each disk \citep[see][]{sierra20}.  This spans the full radial extent of the plots for each disk (with the exception of the inner $\sim$30~au for GM~Aur) demonstrating that the edge of the millimetre dust in each disk is always exterior to edge of the large organic emission.  This is in contrast to, e.g.\ \ce{H2CO}, which appears to be radially extended across all disks studied here \citep[see][]{guzman20}.}  

{We can use the radial column density profiles to estimate the amount of each large organic molecule present in the inner 50~au of each disk (i.e.\ on scales comparable to the Solar System) via conversion to a total surface density and integrating over the corresponding area.}   These values are listed in Table \ref{tab:organics} (left).  While there does not appear to an obvious trend for the gas mass of each large organic molecule between the disks (e.g.\ there appears to be no stellar mass dependence), it appears that \ce{HC3N} is significantly more abundant than either \ce{CH3CN} or $c$-\ce{C3H2} in each disk (by factors of 5--15).  {We can also estimate the total (gas and ice) mass of each of the large organics.  Chemical models of these molecules in disks predict gas-to-ice ratios across a range of values, from 10$^3$ in \citet{Walsh2014} to 10$^{5}$ in \citet{Ruaud2019}, so we conservatively adopt a 1000-to-1 ratio for our conversion.  These values are also shown in Table \ref{tab:organics} (right), expressed as a percentage of \ce{H2O} ice, estimated from predicted abundances in the thermochemical disk models of \citet{zhang20}. This again demonstrates that \ce{HC3N} is most abundant in each disk. AS~209 emerges as a potential outlier, being particularly rich in large organics when compared to the other disks in our small sample by factors of $\sim$30 for \ce{HC3N} and $\sim$10 for both \ce{CH3CN} and $c$-\ce{C3H2}.  However, this may instead be due to a low modelled \ce{H2O} mass, and so observational constraints on this abundance would be required to confirm this.}

\subsection{Chemical origin of the large organics}

While determining the precise chemical origin of the large organic species will require dedicated models for each disk, we can use our empirically-determined quantities to gain insight into general properties. Following \citet{furuya_2014}, we can estimate the typical temperature at which we would expect the large organic molecules to sublimate from icy grain surfaces into the gas phase.  For gas densities in the range of 10$^{6}$--10$^{12}$~cm$^{-3}$, we would expect this sublimation temperature to lie between 80--150~K\footnote{Based on binding energies from KIDA: \url{http://kida.obs.u-bordeaux1.fr}}.  It is therefore interesting that our derived rotational temperatures for \ce{HC3N} and \ce{CH3CN} (where we again note that the assumed rotational temperature of $c$-\ce{C3H2} is tied to the \ce{CO} emission) are well below this value across the full radial extent of each disk.  This suggests that, despite their close association with millimetre dust disk, these molecules cannot originate from direct thermal desorption (with the possible exception of the inner $\sim$20~au of GM~Aur and MWC~480 for \ce{HC3N}, although uncertainties are large in these regions).  Therefore, these species must be released from ices by non-thermal desorption processes (such as those triggered by cosmic rays and/or X-rays), or be formed in the gas phase directly.

\begin{figure*}[!ht]
\centering
\includegraphics[width=\textwidth]{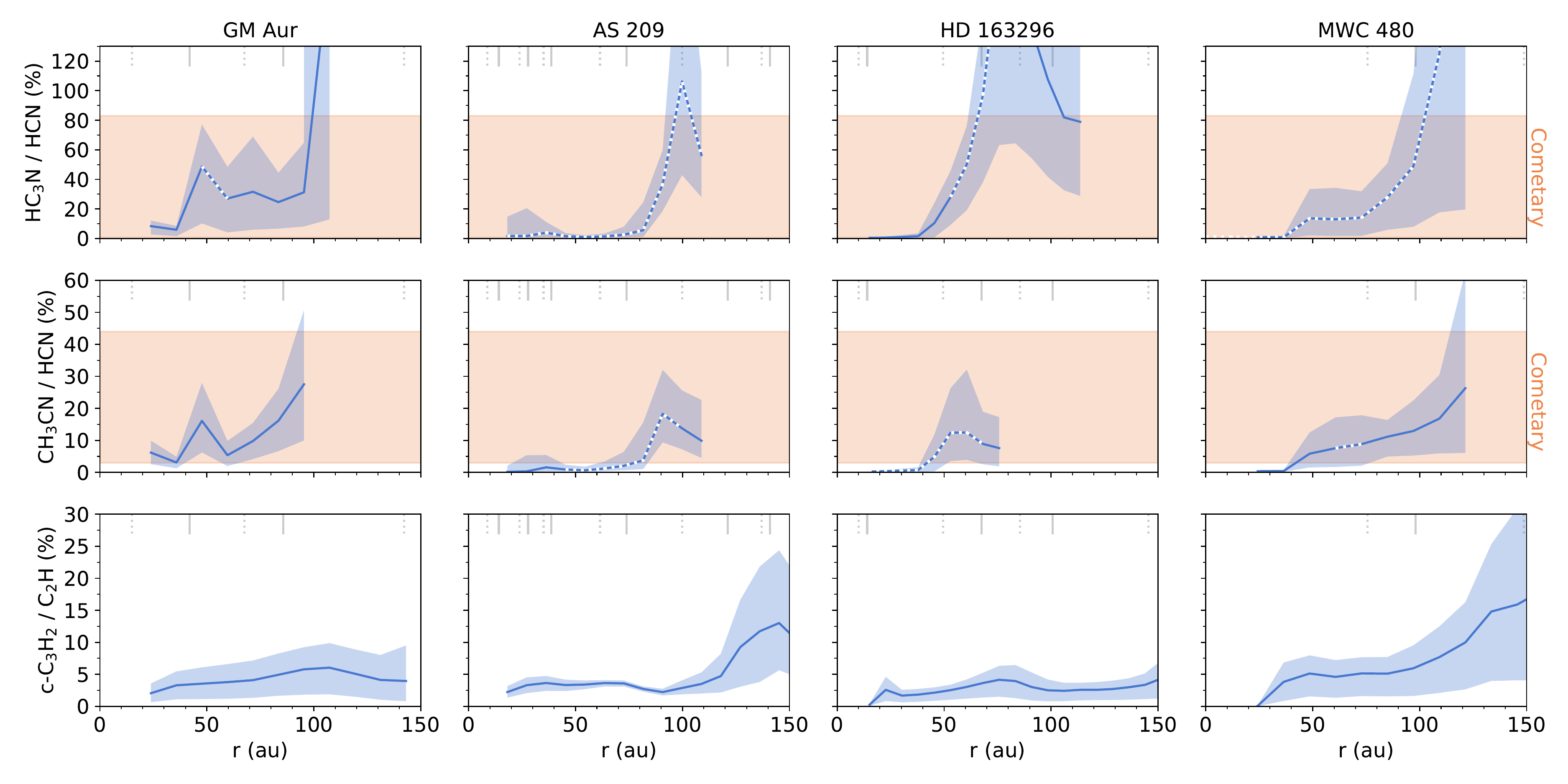}
\caption{Ratio of column densities between the small (\ce{HCN}, \ce{C2H}) and large organic molecules as derived from this work and \citet{guzman20}.  {Regions where our analysis suggests $\tau > 1$ are shown with a dashed line}, and plots have been truncated in regions of low signal-to-noise.  The range of observed values for Solar System comets are shown by the vertical shaded regions \citep[][and references therein]{biver_2019}.  Locations of continuum substructure (e.g. rings and gaps) are marked with grey vertical lines (solid and dotted, respectively).} 
\label{fig:small_organics}
\end{figure*}

We can also compare the column density ratios of simple and complex molecules in each disk to understand the efficiency of conversion from small to large organic species.  Figure \ref{fig:small_organics} shows the radially-resolved ratio of column densities for \ce{HC3N}/\ce{HCN},  \ce{CH3CN}/\ce{HCN} and  $c$-\ce{C3H2}/\ce{C2H} obtained from a comparison of our results with those of \citet{guzman20}.  {We indicate the regions across which our analysis of \ce{HC3N} or \ce{CH3CN} emission suggests $\tau>1$, and thus where the column density ratio may be a lower limit.  We also note that the derivation of \ce{HCN} and \ce{C2H} column density is based on fitting the hyperfine transitions, and is not therefore influenced by the (large) optical depth of the main line component for these molecules.  For the \ce{HC3N}/\ce{HCN} ratio, all disks exhibit lower values in their inner ($\lesssim 50$\,au) regions of approximately 20\% or less.  Beyond radii at which the optical depth drops below 1 in GM~Aur and HD~163296, the measured ratios increase sharply (even given the increase in associated uncertainties).  The \ce{CH3CN}/\ce{HCN} ratio broadly follows a similar pattern to that of the \ce{HC3N}/\ce{HCN}, namely lower values ($\lesssim5$\%) in the inner ($\lesssim 40-80$\,au) region. The \ce{CH3CN}/\ce{HCN} ratio rises gradually in the outer regions of all disks, reaching values of 20--30\% in the outer regions ($>50$\,au) in all disks, where optical depths are below one.   It is interesting to note that the general form of both the \ce{HC3N}/\ce{HCN} and \ce{CH3CN}/\ce{HCN} ratios as a function of radius (e.g. the sharp and shallow rise, respectively) are in agreement with the forward modelling retrieval performed by \citet[][see their figure 9]{Bergner2018}.}

\begin{deluxetable*}{lccc|ccc}
\tablecaption{Large organic reservoir within $50$~au for each disk.\label{tab:organics}}
\tablewidth{0pt}
\tablehead{ 
 & \multicolumn{3}{c}{Gas mass} & \multicolumn{3}{c}{Gas+ice mass w.r.t.\ H$_2$O ice\tablenotemark{$\dagger$}} \\ 
 & \colhead{\ce{HC3N}} & \colhead{\ce{CH3CN}}  & \colhead{$c$-\ce{C3H2}} & \colhead{\ce{HC3N}} & \colhead{\ce{CH3CN}} & \colhead{$c$-\ce{C3H2}} \\
 & \colhead{$10^{21}$~g} & \colhead{$10^{21}$~g} & \colhead{$10^{21}$~g} & \colhead{\% H$_2$O} & \colhead{\% H$_2$O} & \colhead{\% H$_2$O}}
\startdata
IM~Lup 	    & $<0.01$   & $<0.01$   & $<0.01$    &  $<0.01$ &  $<0.01$  &  $<0.01$  \\
GM~Aur 	    & 6.0       & 3.2       & 0.5        &  0.02    & 0.01      & $<0.01$   \\
AS~209 	    & 5.8       & 1.5       & 1.7        &  0.33    & 0.08      & 0.10      \\
HD~163296 	& 6.0       & 2.8       & 1.4      	 &  0.01    & $<0.01$   & $<0.01$   \\
MWC~480 	& 7.8       & 3.2       & 1.1      	 &  0.01    & $<0.01$   & $<0.01$   \\
\enddata
\tablenotetext{\dagger}{Assuming a 1000-to-1 ice-to-gas ratio and \ce{H2O} abundances from \citet{zhang20}}
\end{deluxetable*}

{In contrast to the nitriles, the $c$-\ce{C3H2}/\ce{C2H} ratio is relatively flat across all radii in each of the disks at $\sim5$\%.  Increases of this ratio to levels of 15--20\% are seen in the outer ($\gtrsim100$\,au) regions of AS~209 and MWC~480. {It is highly likely that $c$-\ce{C3H2} is a species that is formed purely in the gas phase in the atmospheres of protoplanetary disks similar to what is found in the interstellar medium \citep[e.g.,][]{Loison2017}.  However, given the denser conditions within protoplanetary disks, a mechanism is required to maintain a source of carbon (chains) in the gas-phase to seed such a chemistry and this could potentially come from the non-thermal desorption of icy hydrocarbon precursors such as \ce{CH4} and \ce{C2H6}.} Hence, this molecule is a counter-example to molecules such as \ce{CH3CN} for which it is known that both gas-phase and ice-phase chemistry are needed to explain its origin in protoplanetary disks \citep[e.g.,][]{Loomis2018a}.
The flat profile of $c$-\ce{C3H2}/\ce{C2H} within 100~au suggests that the gas-phase conversion of simple to complex hydrocarbons is relatively insensitive to radial location in these regions, and occurs at a similar rate in each disk in our sample.}  

{Column density ratios of \ce{HC3N}/\ce{HCN} and \ce{CH3CN}/\ce{HCN} are have been calculated from remote observations of comets (while $c$-\ce{C3H2} has not been detected), allowing us to compare the relative chemical complexity of these disks with the organic material in the Solar System.  \citet{biver_2019} collate these values from numerous observations and find \ce{HC3N}/\ce{HCN} ranges from $\sim 1-80$\%, and \ce{CH3CN}/\ce{HCN} ranges from $\sim 3-45$\%, which we also show on Figure \ref{fig:small_organics}.  The \ce{HC3N}/\ce{HCN} ratio (or lower limit) is consistent with cometary measurements across all of the disks out to $\sim$100~au, and higher beyond this.
The \ce{CH3CN}/\ce{HCN} ratio is mostly consistent with cometary values, but the the inner regions of the AS~209, HD~163296 and MWC~480 disks are somewhat lower than the ratio measured in comets.}  

{The expected formation zone of comets in the Solar System is generally thought to be $\lesssim$40~au \citep[see][]{mumma_2011}.  The large-to-small organic ratios on these scales are difficult to probe in our target disks due to the spatial resolution of our data ($\sim$30--50~au).  However, the general picture that emerges from this comparison is that the outer (50--100~au) regions of all disks are consistent with the composition of cometary material.  In particular, the warmer HD~163296 and MWC~480 disks would likely have comet formation zones at correspondingly larger radii, and so this can be reconciled with a `scaled-up' picture of the Solar System.  If comets were to form in the outer regions of these disks ($\gtrsim$50--100\,au), then they will attain a similar composition of nitriles to those observed in the Solar System.}

\subsection{The (lack of) emission from IM~Lup}

An obvious outlier in our small sample is the IM~Lup disk, which only exhibits tentative detections of \ce{CH3CN} and $c$-\ce{C3H2} and no detection of \ce{HC3N}.  This is in contrast to firm detections of the smaller organic molecules discussed previously \citep{Bergner2019, guzman20}, which demonstrate the precursors of the large organic molecules are at least present in this disk.  IM~Lup is the youngest star-disk system in our sample at 0.2--1.3~Myr old \citep{alcala_2017}.  If these large organic molecules are primarily formed \emph{in situ} within the disk, then IM~Lup may have not had sufficient time to build up a detectable gas-phase reservoir.  This would be in broad agreement with dark cloud chemical models that demonstrate timescales of $10^{5}$ -- $10^{6}$~yrs are required to reach peak abundances for these species \cite[see, e.g.,][]{agundez_2013}.  Alternatively, if the large organic molecules are primarily inherited from the protostellar phase \citep[see, e.g.,][]{Drozdovskaya2016,Drozdovskaya2019,Bianchi2019,Booth2021}, then a short pre-stellar collapse phase would result in a lower abundance of these species in the disk.  

{The physical conditions of the IM~Lup disk might also explain its lack of complex molecular emission.  This disk has been found to be massive, with total disk mass estimates of the order 0.17~M$_{\odot}$ \citep{cleeves_2016}.  This density structure results in an optically-thick region at Band 6 frequencies inside $\sim$50~au, which may be responsible for suppression of emission from CO isotopologues.  A similar suppression of emission from other molecules should also occur, and if the emission originates from closer to the midplane (compared to CO), then this suppression may extend across a larger radial region.  This scenario of flux deficit may explain the relative weakness of the large organic emission in IM~Lup when compared to the other disks in our sample, and is in agreement with detailed studies of line emission from massive disks \citep{evans_2019}.  Disentangling the relative importance of each of these processes on the resultant line emission from large molecules in IM~Lup will require a detailed disk-specific model.}

\subsection{Comparison to similar observational studies}

{\citet{Bergner2018} observed MWC~480 with ALMA and performed a rotational diagram analysis of large organic molecular emission.  For \ce{HC3N}, they found a disk-integrated rotational temperature of $49\pm6$\,K, similar to our derived value of $37.1^{+4.5}_{-2.9}$~K.  However, their column density of $5.8\pm2.8\times10^{12}$\,cm$^{-2}$ is $\sim$13 times lower than the value we calculate here.   This difference stems primarily from the differing spatial resolution of our data, with our 0$\farcs$3 observations able to place tighter constraints on the radial extent of the \ce{HC3N} emission for a disk-integrated analysis.}  For \ce{CH3CN}, \citet{Bergner2018} find a disk-integrated column density of $1.8\pm0.4\times10^{12}$\,cm$^{-2}$ which is similar to our value of $3.5^{+0.2}_{-0.2}\times10^{12}$\,cm$^{-2}$, but a rotational temperature of $73\pm23$\,K, somewhat higher than our $48.6^{+5.7}_{-4.6}$~K.    This could be reconciled if our observations are tracing different emitting layers; \citet{Bergner2018} observed transitions with $E_{\rm u} \sim 150$--250\,K, while our addition of the Band 3 data allows us access transitions down to $E_{\rm u} = 20$\,K. 

\citet{Loomis2018a} observed several transitions of \ce{CH3CN} toward the T Tauri star TW~Hya, performing a rotational diagram analysis across transitions spanning $E_{\rm u} \sim 70$ -- 150\,K.  They derive a disk-integrated column density of $1.45\pm0.2\times10^{12}$\,cm$^{-2}$ and rotational temperature of $32\pm4$\,K.  While the column density is comparable to our values across all disks, the rotational temperature is generally lower (with the exception of AS~209).  This may be indicative of \ce{CH3CN} originating from a cooler region of the TW~Hya disk compared to the objects studied here, or the result of a cooler disk more generally.

\citet{Qi2013} detected $c$-\ce{C3H2} toward HD~163296 using ALMA Science Verification data, finding a single ring structure from $\sim$30--165~au.  They derive a column density of $2.2\pm0.2\times10^{12}$~cm$^{-2}$ at 100\,au, which is very similar to our disk-integrated value.  \citet{Cleeves2021} also recently report a multi-line analysis of $c$-\ce{C3H2} toward TW Hya and through a forward modelling approach find a best-fit column density of 1--$3\times10^{12}$\,cm$^{-2}$ with disk-integrated rotational temperatures for the ortho- and para form of $55\pm13$~K and $43\pm14$~K, respectively.  
Such temperatures are comparable to our assumed values from \ce{^{13}CO} $J=2$--1 between $50-100$\,au in three of our disks (with GM~Aur being $\sim$10~K cooler), and the column density range is comparable to our disk-integrated values.   However, our radially-resolved column density values are a factor of $\sim$10--50 higher.

\subsection{Comparison with disk chemical models}

Our observationally-derived quantities can be compared to chemical models of protoplanetary disks.  While disk-specific models encompassing the full chemistry required to explain the abundances of these molecules are not yet available, more general disk chemical models are still informative.  \citet{Walsh2014} studied the composition of a representative disk around a T Tauri star with a large gas-grain complex chemical network that included both \ce{HC3N} and \ce{CH3CN}.  Their spatial distribution of gaseous \ce{HC3N} and \ce{CH3CN} is characterised by an inner, warm component reaching down to the midplane along with a population of molecules found at higher relative heights ($z/r \gtrsim 0.2$) for larger radii.  While the scales differ, such a morphology is in broad agreement with our derived origin of the line emission in Figure \ref{fig:origin}, though dedicated radiative transfer modelling would be needed to determine those regions of the disk that contribute the most to the emergent flux for the transitions studied here. 

We can also compare our derived molecular abundances to their model via our radially-resolved $N_T$ profiles.  Across our disk sample, peak $N_{T}$(\ce{HC3N}) are between $10^{13}-10^{14}$\,cm$^{-2}$ within $\sim$100~au, but are likely lower limits due to the high optical depth discussed above.  Such values are up to 500 times higher than the \ce{HC3N} column densities seen in the \citealt{Walsh2014} model at similar radii, and even $\sim100$ times higher than those at 10\,au (see their Figure 8 and Table 2).  A less extreme picture emerges when comparing the $N_{T}$(\ce{CH3CN}), where our derived values are $\sim$5--10 times higher than those in the model between 10--100~au.  
Exploring the chemical structure as a function of height at a radius of 300~au, \citet{Walsh2014} also found that gas-phase chemistry only models result in the largest fractional abundance of \ce{HC3N} achieved in the disk atmosphere, whereas models in which grain-surface (ice) chemistry was included achieved the largest fractional abundance of gas-phase \ce{CH3CN}.  This demonstrates the importance of a complete treatment of both gas- and ice-phase reactions in such models.

More recently, \citet{LeGal2019} investigated the effect of changes in C/O ratio in a protoplanetary disk on the resulting abundances of nitrile species including \ce{HC3N} and \ce{CH3CN}.  Their peak column densities for both species (that require ${\rm C}/{\rm O} = 1.0$ and the inclusion of grain-surface formation routes) are approximately $10^{12}$\,cm$^{-2}$, still a factor of 10--100 smaller than the column densities we derive across our disk sample. 
\citet{Wakelam2019} recently explored the impact on vertically-integrated column densities of molecules in protoplanetary disks excluding and including the effects of dust growth and settling on the disk structure and chemistry and also including ice chemistry. 
Of particular interest for our work is the sensitivity of the column density values and distribution of \ce{HC3N}.  
In their fiducial models (no grain growth nor settling), they achieve column densities $\gtrsim 10^{12}$~cm$^{-2}$ only within the inner 20~au, with a typical column density of $\sim 10^{11} - 10^{12}$~cm$^{-2}$ obtained in the outer disk, and increasing with radius. This is similar to the results of \citet{Walsh2014}.  
However, in models where grain growth and settling is included, they obtain an increase of the order of 1 to 2 orders of magnitude in the column density of \ce{HC3N} in the outer disk.  
The model which obtains the highest peak column density is also a model which begins the chemistry with atomic initial abundances, rather than molecular, indicating also the importance of having free carbon available in the gas-phase for the formation of carbon-rich molecules with gas-phase formation pathways such as \ce{HC3N}, despite the bulk gas possessing elemental ratio of C/O~$< 1$.
It is not specifically explained why dust growth and settling boosts the formation of \ce{HC3N}; however, it may to be related to the decrease in altitude of the dust photosphere allowing greater penetration of bond-breaking radiation that then drives a rich gas-phase chemistry. 

The broad picture that emerges from the above comparisons is that current static chemical models generally under-predict the abundance of \ce{HC3N} and \ce{CH3CN} by a significant margin. Only in the case of a gas-grain chemical model that also includes dust growth and settling does the column density, specifically of \ce{HC3N}, approach the values derived here \citep{Wakelam2019}. 
It is also worth to note that the static models presented in \citet{Loomis2018a} are able to well reproduce the radial column density of \ce{CH3CN} in TW~Hya, but only when both gas-phase and ice-phase chemistry are included, along with a high photodesorption rate. 

Beyond an incomplete picture of chemical formation pathways, it is possible that the above disk chemical models are not complete in their description of physical processes that could cause a higher abundance of organic molecules. For example, the radial drift of large ($\gtrsim$mm-sized) grains from the outer regions of the disk can alter the chemical composition of the inner regions, that may manifest in one of two different ways.  If a molecule is formed primarily via grain-surface reactions, then the radial drift of grains will lead to an enhancement of gas-phase abundance within the relevant ice line \citep[see, e.g.,][]{booth_2017}.  However, if the organic molecule is produced more readily in the gas-phase, then radial drift may still play a role.  The redistribution of bulk carriers of major elements via radial drift will alter elemental ratios in the inner disk (for example C/O; see, e.g., \citealt{piso_2015}) that can alter the formation efficiency of molecules such as \ce{HC3N} and \ce{CH3CN} (as demonstrated by \citealt{LeGal2019}).  Quantifying the detailed effects of these dust transport processes on the distribution of large molecules would require coupled models that include radial drift of dust and gas-grain chemical kinetics.  While such models employing small chemical networks are beginning to emerge \citep[see, e.g.,][]{booth_2019}, they have not yet been expanded to include the reactions required to track the chemical evolution of larger molecules. 

In addition to the radial motions of solids, vertical motions of gas and dust could also act to alter the abundances of molecules in a disk.  \citet{Semenov_2011} modelled the chemical evolution of a disk including turbulent transport of gas and dust in the form of turbulent diffusion.  They find that the column densities of species we study here -- \ce{HC3N}, \ce{CH3CN} and $c$-\ce{C3H2} -- can be enhanced by up to factors of 10--20 in the case of a fast mixing scenario compared with a static disk.  
This enhancement occurs when grains with icy mantles are transported from the disk midplane to warmer vertical layers, causing heavy radical species in the ices to become more mobile.  This leads to the production of more complex molecules on the grain surfaces, that are then released into the gas.    
Models presented in \citet[][]{Oberg2015} to aid the interpretation of the first detections of \ce{HC3N} and \ce{CH3CN} in MWC~480 also explored the impact of vertical mixing driven by turbulence on the column densities of both species.  While high levels of turbulence are in tension with current measurements in the line-emitting regions of protoplanetary disks ($\alpha \lesssim 10^{-3}$; e.g. \citealt{Flaherty20}) and observations of well-settled millimetre dust \citep{villenave_2020}, it was shown that even models with low levels of turbulent mixing better reproduced the column densities derived from observations.  It is also possible that mixing is driven via local mechanisms such as meridional flows generated by the accretion of embedded planets \citep{teague_2019}; however, the impact of this is yet to be investigated in chemical models of protoplanetary disks.

\subsection{Future Outlook}

Despite the impressive diagnostic power of the MAPS observations we present here, it is clear that we are operating at the limits of the data.  For the weakest transitions (e.g.\ \ce{CH3CN} 6--5) we are primarily limited by sensitivity, and so future observational campaigns aiming to characterise larger molecules in disks should be designed with this in mind.  We have shown that the combination of observations for transitions spanning a large range of upper state energy are particularly powerful, and so this should also be maximised.  Further characterisation of the gas substructure for the brighter transitions (e.g.\ $c$-\ce{C3H2} $7_{07}$--$6_{16}$) would benefit from deeper, higher angular resolution studies, which may elucidate any (anti-) correlations with the millimetre dust substructures that are so well characterised in these disks.  Our findings also demonstrate that the current state-of-the-art for protoplanetary disk chemical models may require extension if they are to explain the latest observations.  Whether this extension requires the addition of further physical or chemical processes remains to be seen.  Nevertheless, it is clear that high angular resolution and sensitivity observations are essential in order to further our understanding of the complex chemistry in disks.

\section{Summary}

In this work we have analysed observations of the large organic molecules \ce{HC3N}, \ce{CH3CN} and $c$-\ce{C3H2} toward five protoplanetary disks as part of the ALMA MAPS Large Program.  We summarise our findings below:

\begin{itemize}

\item We robustly detect multiple transitions of \ce{HC3N}, \ce{CH3CN} and $c$-\ce{C3H2} in AS~209, GM~Aur, HD~163296 and MWC~480.  For IM~Lup, we only tentatively detect single transitions of $c$-\ce{C3H2} and \ce{CH3CN}.

\item There appears to be a weak relationship between millimetre dust morphology and the morphology of line emission from these molecules.  There are no disk-wide trends between, for example, the depletion of millimetre dust and the depletion of any molecule in the gas (with the possible exception of $c$-\ce{C3H2} in AS~209, HD~163296 and MWC~480).  

\item Disk-integrated column densities and temperatures are broadly consistent with previous observational studies, where available.  However, the high angular resolution of our observations allows us to investigate radially-resolved properties, revealing a significant increase in molecular column density for \ce{HC3N} and \ce{CH3CN} at small radii.  Emission from these molecules appears to be optically thick within 50--100\,au in all disks, suggesting that these higher column densities are likely lower limits.

\item The emission in all disks is compact ($\lesssim$100~au) and close to the extent of the millimetre dust disk, suggesting that the ice reservoirs hosted on grains may play a role in the formation of these species.  However, the derived rotational temperatures are below the expected sublimation temperatures for each of the molecules, ruling out thermal desorption as an origin. 

\item Comparison with existing disk chemical models shows that static models cannot generally reproduce the high column densities we observe.  In contrast, chemical models that include dynamic processes such as radial drift and vertical mixing could explain larger column densities that we observe in these disks.

\item We approximate the two-dimensional $(r,z)$ locations where we expect the molecules to emit, finding that \ce{CH3CN} originates from close to the midplane in all disks ($z/r \lesssim 0.1$), while \ce{HC3N} originates in higher layers ($z/r \sim 0.1$--0.4). This is consistent with distributions predicted by disk chemical models.  {In addition, we find that emission occurs at higher $z/r$ in the disks around T Tauri stars when compared with those around the Herbig stars.}

\item We find good agreement between relative abundances of simple and complex nitrile species in {the outer regions of} each disk, and with remote observations of Solar System comets.  The conversion efficiency of small to large hydrocarbons appears to be low (5--10\%) and generally insensitive to radial location within the disks.

\end{itemize}

We have demonstrated that four of the protoplanetary disks studied here -- GM~Aur, AS~209, HD~163296, and MWC~480 -- all contain significant reservoirs of the large organic molecules \ce{HC3N}, \ce{CH3CN} and $c$-\ce{C3H2} on scales comparable with the extent of their millimetre dust disks.   These dust disks all host rings and gaps, and HD~163296 and MWC~480 exhibit deviations from Keplerian velocities in their CO gas emission \citep{teague20}  In many cases, these phenomena can be readily explained by the presence of forming planets.  Our analysis also shows that these molecules can emit from close to the disk midplane {(particularly for the disks around the Herbig stars)}, where the majority of planet formation processes operate.  Our results are therefore consistent with a picture in which the precursors to biologically-relevant molecules are abundant in the raw material available for planet formation in protoplanetary disks, and that this material can have a similar composition to that within our own Solar System.

\acknowledgments

{We are grateful to the anonymous referee for a constructive report that improved the clarity of the manuscript.}  This paper makes use of the following ALMA data: ADS/JAO.ALMA\#2018.1.01055.L. ALMA is a partnership of ESO (representing its member states), NSF (USA) and NINS (Japan), together with NRC (Canada), MOST and ASIAA (Taiwan), and KASI (Republic of Korea), in cooperation with the Republic of Chile. The Joint ALMA Observatory is operated by ESO, AUI/NRAO and NAOJ. The National Radio Astronomy Observatory is a facility of the National Science Foundation operated under cooperative agreement by Associated Universities, Inc.

J.D.I. acknowledges support from the Science and Technology Facilities Council of the United Kingdom (STFC) under ST/T000287/1. 
C.W. acknowledges financial support from the University of Leeds, STFC and UKRI (grant numbers ST/R000549/1, ST/T000287/1, MR/T040726/1). 
A.S.B acknowledges the studentship funded by the Science and Technology Facilities Council of the United Kingdom (STFC).
Y.A. acknowledges support by NAOJ ALMA Scientific Research Grant Code 2019-13B, and Grant-in-Aid for Scientific Research 18H05222 and 20H05847. 
S.M.A. and J.H. acknowledge funding support from the National Aeronautics and Space Administration under Grant No. 17-XRP17 2-0012 issued through the Exoplanets Research Program. 
J.B. acknowledges support by NASA through the NASA Hubble Fellowship grant \#HST-HF2-51427.001-A awarded  by  the  Space  Telescope  Science  Institute,  which  is  operated  by  the  Association  of  Universities  for  Research  in  Astronomy, Incorporated, under NASA contract NAS5-26555.
E.A.B. and A.D.B. acknowledge support from NSF AAG Grant \#1907653.
J.B.B. acknowledges support from NASA through the NASA Hubble Fellowship grant \#HST-HF2-51429.001-A, awarded by the Space Telescope Science Institute, which is operated by the Association of Universities for Research in Astronomy, Inc., for NASA, under contract NAS5-26555.
G.C. is supported by NAOJ ALMA Scientific Research Grant Code 2019-13B. 
L.I.C. gratefully acknowledges support from the David and Lucille Packard Foundation and Johnson \& Johnson's WiSTEM2D Program. 
I.C. was supported by NASA through the NASA Hubble Fellowship grant HST-HF2-51405.001-A awarded by the Space Telescope Science Institute, which is operated by the Association of Universities for Research in Astronomy, Inc., for NASA, under contract NAS5-26555. 
V.V.G. acknowledges support from FONDECYT Iniciaci\'on 11180904 and ANID project Basal AFB-170002.
J. H. acknowledges support for this work provided by NASA through the NASA Hubble Fellowship grant \#HST-HF2-51460.001-A awarded by the Space Telescope Science Institute, which is operated by the Association of Universities for Research in Astronomy, Inc., for NASA, under contract NAS5-26555. 
C.J.L. acknowledges funding from the National Science Foundation Graduate Research Fellowship under Grant DGE1745303. 
R.L.G. acknowledges support from a CNES fellowship grant.
F.M. acknowledges support from ANR of France under contract ANR-16-CE31-0013 (Planet Forming Disks) and ANR-15-IDEX-02 (through CDP ``Origins of Life").
H.N. acknowledges support by NAOJ ALMA Scientific Research Grant Code 2018-10B and Grant-in-Aid for Scientific Research 18H05441.
K.I.\"O. acknowledges support from the Simons Foundation (SCOL \#321183) and an NSF AAG Grant (\#1907653). 
K.R.S. acknowledges the support of NASA through Hubble Fellowship Program grant HST-HF2-51419.001, awarded by the Space Telescope Science Institute,which is operated by the Association of Universities for Research in Astronomy, Inc., for NASA, under contract NAS5-26555. 
R.T. acknowledges support from the Smithsonian Institution as a Submillimeter Array (SMA) Fellow. 
T.T. is supported by JSPS KAKENHI Grant Numbers JP17K14244 and JP20K04017. 
Y.Y. is supported by IGPEES, WINGS Program, the University of Tokyo. 
K.Z. acknowledges the support of the Office of the Vice Chancellor for Research and Graduate Education at the University of Wisconsin – Madison with funding from the Wisconsin Alumni Research Foundation, and the support of NASA through Hubble Fellowship grant HST-HF2-51401.001. awarded by the Space Telescope Science Institute, which is operated by the Association of Universities for Research in Astronomy, Inc., for NASA, under contract NAS5-26555. 

\vspace{5mm}
\facilities{ALMA}

\software{
CASA \citep{mcmullin07},
Astropy \citep{astropy:2013}, 
Matplotlib \citep{Hunter:2007},
NumPy \citep{harris2020array},
emcee \citep{emcee_2013},
bettermoments \citep{teague_bettermoments},
GoFish \citep{teague_gofish},
VISIBLE \citep{Loomis2018b}.
}

\appendix

\section{Gallery of integrated intensity (zeroth moment) maps}
\label{sec:mom0}

\begin{figure*}[!ht]
\centering
\includegraphics[width=\textwidth,trim=0 0cm 0 0cm, clip]{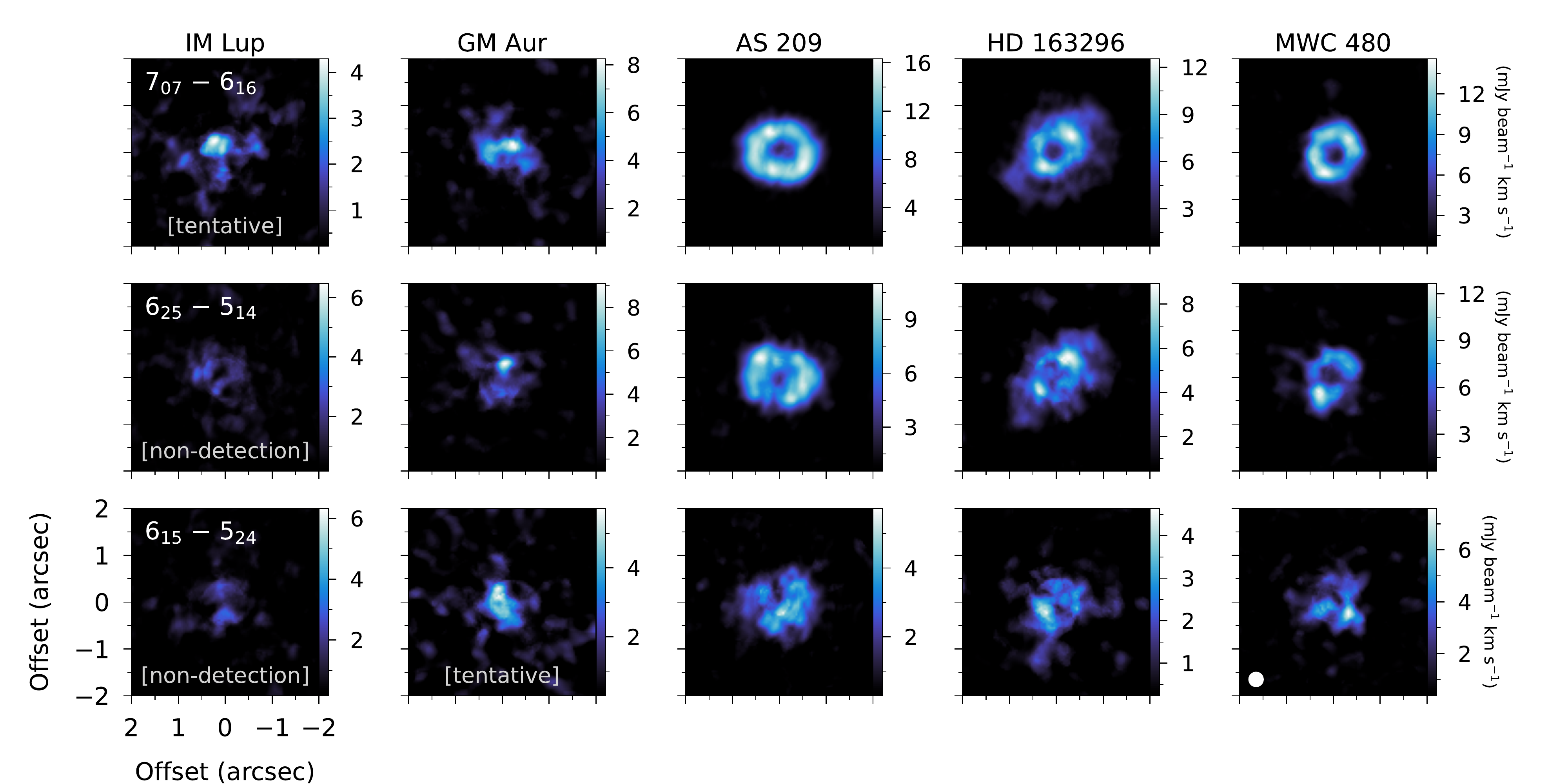}
\caption{Integrated intensity (zeroth moment) maps for transitions of $c$-\ce{C3H2}.}
\label{fig:mom0_c-C3H2}
\end{figure*}

\begin{figure*}[!ht]
\centering
\includegraphics[width=\textwidth,trim=0 0cm 0 0cm, clip]{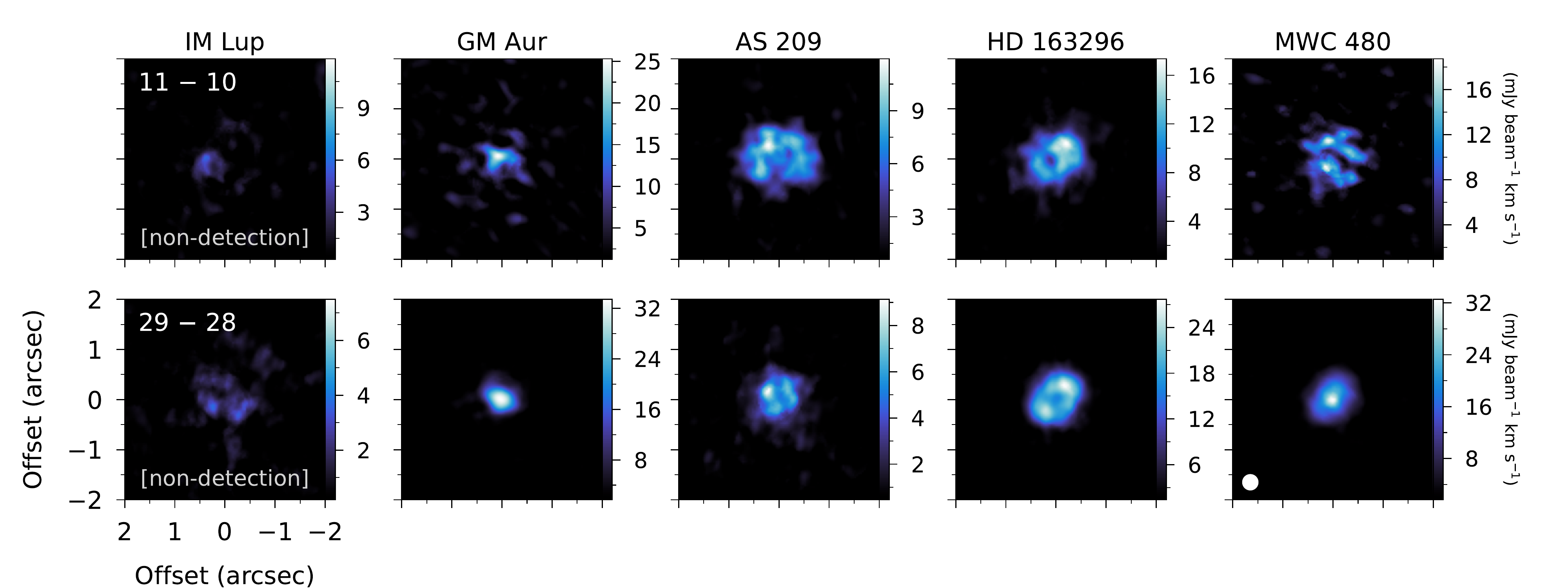}
\caption{Integrated intensity (zeroth moment) maps for transitions of \ce{HC3N}.}
\label{fig:mom0_HC3N}
\end{figure*}

\begin{figure*}[!ht]
\centering
\includegraphics[width=\textwidth,trim=0 0cm 0 0cm, clip]{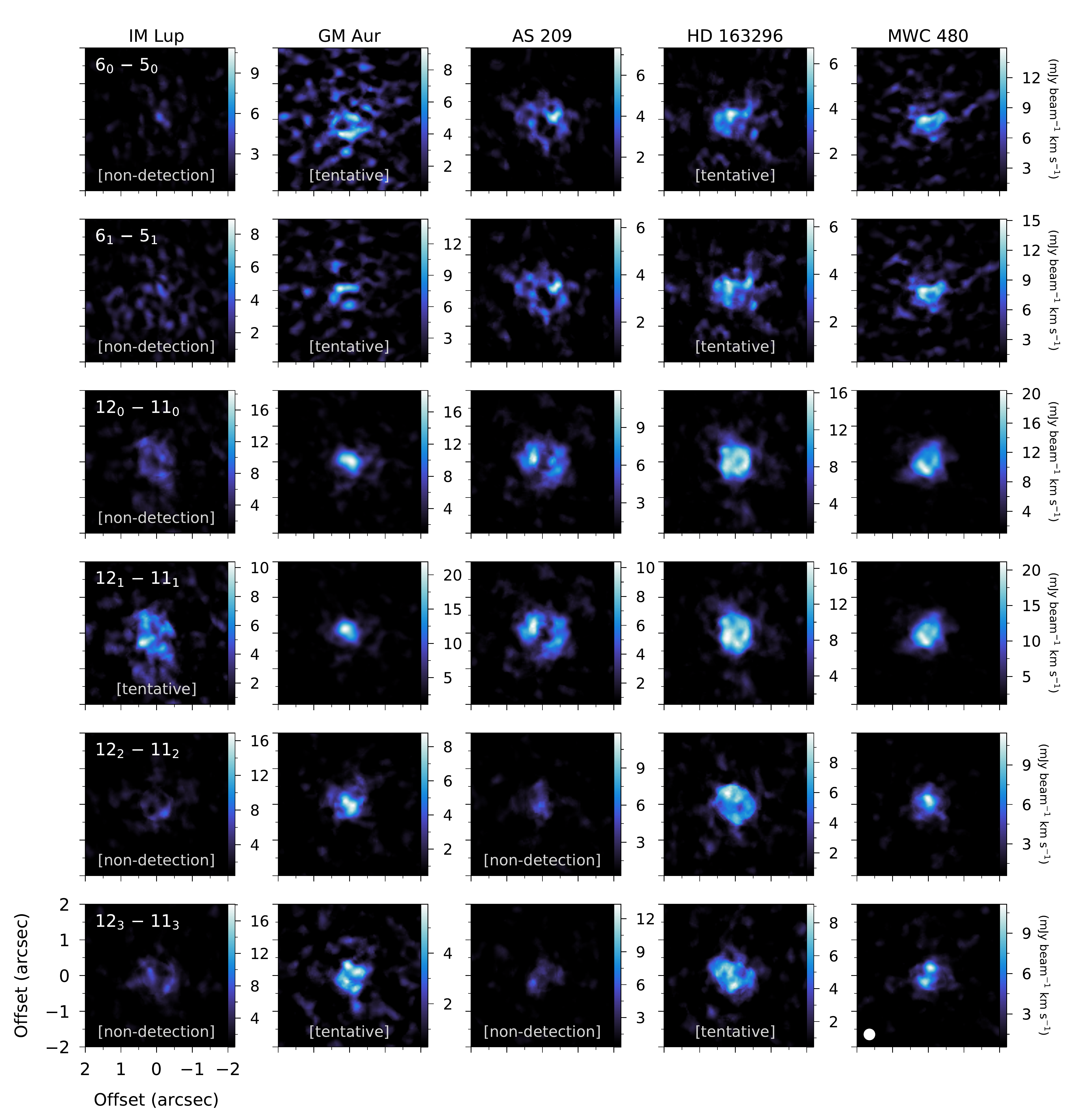}
\caption{Integrated intensity (zeroth moment) maps for transitions of \ce{CH3CN}.}
\label{fig:mom0_CH3CN}
\end{figure*}

Figures \ref{fig:mom0_c-C3H2}, \ref{fig:mom0_HC3N} and \ref{fig:mom0_CH3CN} show the integrated intensity maps for all disks and transitions studied in this work.

\section{Molecular data}
\label{sec:molec}

Table \ref{tab:molecular} details the molecular spectroscopic data and constants that have been used throughout this work.

\begin{deluxetable}{llcccc}
\tablecaption{List of transitions and associated molecular spectroscopic data, taken from the Cologne Database for Molecular Spectroscopy\tablenotemark{a} (CDMS).\label{tab:molecular}}
\tablewidth{0pt}
\tablehead{\colhead{Transition} & \colhead{Quantum Numbers} & \colhead{Frequency} & \colhead{$E_\mathrm{u}$} & \colhead{log$_{10}(A_{\rm ij})$} & \colhead{$g_\mathrm{u}$} \\
  &  & \colhead{(GHz)} & \colhead{(K)} & }
\startdata
\sidehead{\ce{HC3N}}
11--10                &  $J=11-10$       & 100.0763920   & 28.8      & $-4.1096$       & 23\\
29--28                &  $J=29-28$       & 263.7923080   & 189.9     & $-2.8349$       & 59\\
\sidehead{\ce{CH3CN}}
 $6_5$--$5_5$        & $J=6-5$, $K=5$         & 110.3303454   & 197.1     & $-4.4697$       & 26\\
 $6_4$--$5_4$        & $J=6-5$, $K=4$         & 110.3494707   & 132.8     & $-4.2098$       & 26\\
 $6_3$--$5_3$        & $J=6-5$, $K=3$         & 110.3643540   & 82.8      & $-4.0792$       & 52\\
 $6_2$--$5_2$        & $J=6-5$, $K=2$         & 110.3749894   & 47.1      & $-4.0054$       & 26\\
 $6_1$--$5_1$        & $J=6-5$, $K=1$         & 110.3813723   & 25.7      & $-3.9664$       & 26\\
 $6_0$--$5_0$        & $J=6-5$, $K=0$         & 110.3835002   & 18.5      & $-3.9542$       & 26\\
 $12_4$--$11_4$      & $J=12-11$, $K=4$         & 220.6792874   & 183.1     & $-3.0857$       & 50\\
 $12_3$--$11_3$      & $J=12-11$, $K=3$         & 220.7090174   & 133.2     & $-3.0624$       & 100\\
 $12_2$--$11_2$      & $J=12-11$, $K=2$         & 220.7302611   & 97.4      & $-3.0465$       & 50\\
 $12_1$--$11_1$      & $J=12-11$, $K=1$         & 220.7430111   & 76.0      & $-3.0372$       & 50\\
 $12_0$--$11_0$      & $J=12-11$, $K=0$         & 220.7472617   & 68.9      & $-3.0342$       & 50\\
\sidehead{$c$-\ce{C3H2}}
 $7_{07}$--$6_{16}$\tablenotemark{$\ddag$}  & $J=7-6$, $K_{a}=0-1$, $K_{c}=7-6$         & 251.3143670   & 50.7      & $-3.0704$       & 45\\ 
 $6_{15}$--$5_{24}$  & $J=6-5$, $K_{a}=1-2$, $K_{c}=5-4$         & 251.5087085   & 47.5      & $-3.1706$       & 13\\
 $6_{25}$--$5_{14}$  & $J=6-5$, $K_{a}=2-1$, $K_{c}=5-4$         & 251.5273110   & 47.5      & $-3.1708$       & 39\\
\enddata
\tablenotetext{a}{\url{https://cdms.astro.uni-koeln.de/cdms/portal/}}
\tablenotetext{$\ddag$}{Includes blended para transition ($7_{17}$--$6_{06}$) at the same frequency, see Section \ref{res:ncolandtrot}.}
\end{deluxetable}

\section{Disk Temperature Structures}
\label{sec:temps}

Table \ref{tab:temps} details the disk temperature structures that have been adopted during the rotational analysis of the $c$-\ce{C3H2} emission.  For further details, see \citet{law20_vert}.

\begin{deluxetable*}{lccccc}
\tablecaption{Summary of parameters used for the temperature structures of the disks (see \citealt{law20_vert}).\label{tab:temps}}
\tablewidth{0pt}
\tablehead{\colhead{}   & \colhead{IM Lup}  & \colhead{GM Aur}  & \colhead{AS 209}  & \colhead{HD 163296}   & \colhead{MWC 480}}
\startdata
\sidehead{$T(r)$ parameters (\ce{^{13}CO} $J=2-1$):}
\phn$r_{\rm{fit, in}}$ (au) &  145              & 50                   &  125                 & 50                 & 100                  \\
\phn$r_{\rm{fit, out}}$ (au)&  339              & 314                  &  163                 & 356                & 388                  \\
\phn$T_{100}$ (K)           &  30~$\pm$~0.6     & 22~$\pm$~0.2         &  28~$\pm$~1.3        & 31~$\pm$~0.2       & 42~$\pm$~0.9         \\
\phn$q$                     &  0.323~$\pm$~0.03 & 0.260~$\pm$~0.01     &  0.804~$\pm$~0.13    & 0.367~$\pm$~0.01   & 0.598~$\pm$~0.03     \\
\sidehead{$T(r,z)$ parameters:}
\phn$T_{\rm{atm},0}$ (K)    &  36$^{+0.1}_{-0.1}$         & 48$^{+0.3}_{-0.3}$        & 37$^{+0.2}_{-0.2}$        & 63$^{+0.2}_{-0.2}$          & 69$^{+0.2}_{-0.2}$          \\
\phn$T_{\rm{mid},0}$ (K)    &  25$^{+0.1}_{-0.1}$         & 20$^{+0.2}_{-0.2}$        & 25$^{+0.2}_{-0.2}$        & 24$^{+0.1}_{-0.1}$          & 27$^{+0.2}_{-0.2}$          \\   
%\phn$q_{\rm{mid}}$          &  $-$0.02$^{+0.01}_{-0.01}$  & $-$0.01$^{+0.01}_{-0.01}$ & $-$0.18$^{+0.01}_{-0.01}$ & $-$0.18$^{+0.004}_{-0.004}$ & $-$0.23$^{+0.01}_{-0.01}$   \\
%\phn$z_0$ (au)              &  3$^{+0.1}_{-0.1}$          & 13$^{+0.2}_{-0.2}$        & 5$^{+0.2}_{-0.2}$         & 9$^{+0.1}_{-0.1}$           & 7$^{+0.1}_{-0.1}$           \\
%\phn$\alpha$                &  4.91$^{+0.17}_{-0.16}$     & 2.57$^{+0.03}_{-0.03}$    & 3.31$^{+0.12}_{-0.11}$    & 3.01$^{+0.02}_{-0.02}$      & 2.78$^{+0.02}_{-0.02}$      \\
%\phn$\beta$                 &  2.07$^{+0.02}_{-0.02}$     & 0.54$^{+0.01}_{-0.01}$    & 0.02$^{+0.02}_{-0.02}$    & 0.42$^{+0.004}_{-0.004}$    & $-$0.05$^{+0.01}_{-0.01}$   \\
\enddata
\end{deluxetable*}

\bibliography{MAPS}{}
\bibliographystyle{aasjournal}

\end{document}

%% file: sigma_flux_table_R1.tex
\begin{deluxetable*}{lccccccccccc}
\tablecaption{Peak matched filter response ($\sigma_{\mathrm{f}}$) and disk-integrated fluxes ($S_{\nu} \Delta v$) for each transition and disk. \label{tab:flux}}
\tabletypesize{\footnotesize}
\tablewidth{0pt}
\tablehead{ 
&  & \multicolumn{2}{c}{IM~Lup} & \multicolumn{2}{c}{GM~Aur} & \multicolumn{2}{c}{AS~209} & \multicolumn{2}{c}{HD~163296} & \multicolumn{2}{c}{MWC~480} \\
\colhead{Transition} & \colhead{Frequency} & \colhead{$\sigma_{\mathrm{f}}$} & \colhead{$S_{\nu} \Delta v$} & \colhead{$\sigma_{\mathrm{f}}$} & \colhead{$S_{\nu} \Delta v$} & \colhead{$\sigma_{\mathrm{f}}$} & \colhead{$S_{\nu} \Delta v$} & \colhead{$\sigma_{\mathrm{f}}$} & \colhead{$S_{\nu} \Delta v$} & \colhead{$\sigma_{\mathrm{f}}$} & \colhead{$S_{\nu} \Delta v$} \\
 & \colhead{(GHz)} &  & \colhead{(mJy\,km\,s$^{-1}$)} &  & \colhead{(mJy\,km\,s$^{-1}$)} &  & \colhead{(mJy\,km\,s$^{-1}$)} &  & \colhead{(mJy\,km\,s$^{-1}$)} &  & \colhead{(mJy\,km\,s$^{-1}$)}   
 }
\startdata                                                                              
\sidehead{\ce{HC3N}}                        
 11--10                & 100.0763920   & $<3$  & $<11.5$                   & 14.7  & $31.3 \pm 5.5$    & 27.6 & $118.0 \pm 6.6$       & 32.8  & $128.8 \pm  5.6$                & 26.9  & $74.1 \pm  5.8$     \\
 29--28                & 263.7923080   & $<3$  & $<4.3$                    & 11.1  & $87.7 \pm 2.3$    & 9.9  & $38.5  \pm 2.3$       & 24.4  & $196.9 \pm  3.0$                & 19.0  & $124.2 \pm 3.1$      \\
\sidehead{\ce{CH3CN}}
 $6_5$--$5_5$          & 110.3303454   & $<3$  & $<3.9$                    & $<3$  & $<2.9$            & $<3$ & $<5.3$                & $<3$  & $<2.7$                          & $<3$  & $<7.6$             \\
 $6_4$--$5_4$          & 110.3494707   & $<3$  & $<3.6$                    & $<3$  & $<2.6$            & $<3$ & $<5.1$                & $<3$  & $<2.5$                          & $<3$  & $<8.2$             \\
 $6_3$--$5_3$          & 110.3643540   & $<3$  & $<3.6$                    & $<3$  & $<2.7$            & $<3$ & $<5.4$                & $<3$  & $<2.7$                          & $<3$  & $<8.8$             \\
 $6_2$--$5_2$          & 110.3749894   & $<3$  & $<3.5$                    & 3.1   & $1.1 \pm 0.9$     & $<3$ & $<5.7$                & $<3$  & $<2.8$                          & $<3$  & $<9.2$             \\
 $6_1$--$5_1$\tablenotemark{$\dag$}          & 110.3813723   & $<3$  & $<3.6$                    & 3.4   & $5.3 \pm 1.0$     & 5.5  & $12.1 \pm 1.8$        & 4.1   & $5.6 \pm 0.9$                   & 6.7   & $16.6 \pm 2.8$     \\
 $6_0$--$5_0$\tablenotemark{$\dag$}          & 110.3835002   & $<3$  & $<3.6$                    & 3.2   & $3.0 \pm 0.9$     & 6.1  & $8.6 \pm 1.7$         & 3.0   & $7.8 \pm 0.9$                   & 7.7   & $8.9  \pm 2.7$     \\
 $12_3$--$11_3$        & 220.7090174   & $<3$  & $<3.7$                    & 4.8   & $5.9 \pm 0.3$     & $<3$ & $<4.3$                & 3.2   & $8.2 \pm 0.8$                   & 9.1   & $22.8 \pm 1.3$    \\
 $12_2$--$11_2$        & 220.7302611   & $<3$  & $<3.5$                    & 7.9   & $9.3 \pm 0.4$     & $<3$ & $<4.1$                & 6.5   & $12.5 \pm 0.8$                  & 8.0   & $26.1 \pm 1.3$    \\
 $12_1$--$11_1$\tablenotemark{$\dag$}        & 220.7430111   & 3.7   & $12.1 \pm  1.2$           & 12.6  & $12.8 \pm 0.4$    & 7.4  & $15.3 \pm 1.5$        & 8.2   & $22.8 \pm 0.9$                  & 14.1  & $31.4 \pm 1.4$    \\
 $12_0$--$11_0$\tablenotemark{$\dag$}        & 220.7472617   & $<3$  & $<3.6$                    & 10.9  & $14.2 \pm 0.4$    & 9.2  & $24.3 \pm 1.4$        & 7.9   & $22.3 \pm 0.8$                  & 16.4  & $44.3 \pm 1.2$    \\
\sidehead{\ce{c-C3H2}} 
 $7_{07}$--$6_{16}$\tablenotemark{$\ddag$}    & 251.3143670   & 4.2   & $6.1 \pm 1.4$            & 8.4   & $27.4 \pm 1.6$    & 52.2 & $228.1 \pm 5.8$       & 51.1  & $146.8 \pm 2.8$                 & 25.0  & $119.9 \pm 3.2$   \\
 $6_{15}$--$5_{24}$    & 251.5087085   & $<3$  & $<4.4$                   & $<3$  & $<4.7$            & 11.2 & $43.9  \pm 2.4$       & 9.8   & $28.4  \pm 1.9$                 & 5.4   & $20.9  \pm 2.7$    \\
 $6_{25}$--$5_{14}$    & 251.5273110   & $<3$  & $<4.3$                   & 7.2   & $11.5 \pm 1.7$    & 31.9 & $133.3 \pm 4.0$       & 28.4  & $91.4  \pm 2.5$                 & 14.8  & $72.2  \pm 2.7$    \\
\enddata
\tablenotetext{$\dag$}{\ce{CH3CN} $K=0$ and $K=1$ are blended, fluxes are measured following the method outlined in Section \ref{sec:gofish}}
\tablenotetext{$\ddag$}{Includes blended para transition ($7_{17}$--$6_{06}$) at the same frequency, see Section \ref{res:ncolandtrot}.}
\end{deluxetable*}

%% file: disk_int_fits_R1.tex
\begin{deluxetable*}{lccccccccc}
\def\arraystretch{1.5}
\tablecaption{Median values of column density (N$_{\rm T}$) and rotational temperature (T$_{\rm rot}$) from the posterior probability distribution of the MCMC fitting procedure to the disk-integrated rotational diagrams (where uncertainties correspond to the 16$^{\rm th}$--84$^{\rm th}$ percentile) along with the maximum optical depth ($\tau_{\rm max}$) calculated for each transition. \label{tab:disk_rot}}
\tablewidth{0pt}
\tablehead{
\colhead{}  & \multicolumn3c{\ce{HC3N}}                                         & \multicolumn3c{\ce{CH3CN}}                                        & \multicolumn3c{\ce{c-C3H2}}   \\
\colhead{}  & N$_{\rm T}$  & T$_{\rm rot}$ & $\tau_{\rm max}$                   & N$_{\rm T}$ & T$_{\rm rot}$ & $\tau_{\rm max}$                    & N$_{\rm T}$ & T$_{\rm rot}$ & $\tau_{\rm max}$ \\
\colhead{}  & (cm$^{-2}$)  & (K)           &                                    & (cm$^{-2}$) & (K)           &                                     & (cm$^{-2}$) & (K)           &         }           
\startdata
IM Lup      &  $<5.5\times10^{12}$ & \nodata & \nodata                          & $<6.5\times10^{11}$ & \nodata & \nodata                            & $<7.3\times10^{11}$ & \nodata & \nodata   \\ 
GM Aur      &  $1.9^{+0.4}_{-0.4}\times10^{13}$ & $54.4^{+5.8}_{-4.3}$  & 0.40   & $2.1^{+0.2}_{-0.1}\times10^{12}$ & $40.8^{+3.2}_{-2.8}$    & 0.04 & $1.1^{+0.1}_{-0.1}\times10^{12}$  & $20$--$48$         & 0.04   \\ 
AS 209      &  $2.9^{+0.5}_{-0.5}\times10^{13}$ & $31.1^{+1.2}_{-1.0}$  & 1.85   & $1.7^{+0.2}_{-0.2}\times10^{12}$ & $24.8^{+2.3}_{-1.8}$    & 0.11 & $7.0^{+0.3}_{-0.2}\times10^{12}$  & $25$--$37$         & 0.24    \\   
HD 163296   &  $7.3^{+2.5}_{-1.9}\times10^{13}$ & $36.9^{+3.1}_{-2.2}$  & 2.62   & $2.3^{+0.2}_{-0.2}\times10^{12}$ & $35.1^{+2.3}_{-1.9}$    & 0.07 & $2.1^{+0.1}_{-0.1}\times10^{12}$  & $24$--$63$         & 0.03   \\ 
MWC 480     &  $7.8^{+3.9}_{-2.7}\times10^{13}$ & $37.1^{+4.5}_{-2.9}$  & 2.83   & $3.5^{+0.2}_{-0.2}\times10^{12}$ & $48.6^{+5.7}_{-4.6}$    & 0.06 & $5.7^{+0.3}_{-0.2}\times10^{12}$  & $27$--$69$         & 0.14     \\ 
\enddata
\end{deluxetable*}